\documentclass[superscriptaddress,twocolumn,secnumarabic,amssymb,amsmath,nobibnotes,aps,prd,nofootinbib]{revtex4}%
\usepackage{graphicx}
\usepackage{epsf}
\usepackage{bm}
\usepackage{amsmath}
\usepackage{amsfonts}
\usepackage{amssymb}
\usepackage{epstopdf}
\usepackage{natbib}
\usepackage{color}%
\providecommand{\U}[1]{\protect\rule{.1in}{.1in}}
\newcommand{\be}{\begin{equation}}
\newcommand{\ee}{\end{equation}}

\newcommand{\mincir}{\raise
	-3.truept\hbox{\rlap{\hbox{$\sim$}}\raise4.truept\hbox{$<$}\ }}
\newcommand{\magcir}{\raise
	-3.truept\hbox{\rlap{\hbox{$\sim$}}\raise4.truept\hbox{$>$}\ }}

\usepackage{color}

\begin{document}

\title{Observational constraints of a new unified dark fluid and the $H_0$ tension}

\author{Weiqiang Yang}
\email{d11102004@163.com}
\affiliation{Department of Physics, Liaoning Normal University, Dalian, 116029, P. R.
China}

\author{Supriya Pan}
\email{supriya.maths@presiuniv.ac.in}
\affiliation{Department of Mathematics, Presidency University, 86/1 College Street, Kolkata 700073, India}

\author{Andronikos Paliathanasis}
\email{anpaliat@phys.uoa.gr}
\affiliation{Institute of Systems Science, Durban University of Technology, PO Box 1334,
Durban 4000, Republic of South Africa}

\author{Subir Ghosh}
\email{subirghosh20@gmail.com}
\affiliation{Physics and Applied Mathematics Unit, Indian Statistical
Institute, 203 Barrackpore Trunk Road, Kolkata-700108, India}

\author{Yabo Wu} 
\email{ybwu61@163.com}
\affiliation{Department of Physics, Liaoning Normal University, Dalian, 116029, P. R. China}


\begin{abstract}
Unified cosmological models  have received a lot of  attention in astrophysics community for explaining both the dark matter and dark energy evolution. The Chaplygin cosmologies, a well known name in this group have been investigated matched  with observations from different sources. Obviously, Chaplygin cosmologies have to obey restrictions in order to be consistent with the observational data.  As a consequence, alternative unified models, differing from Chaplygin model,  are of special interest. In the present work we consider a specific example of  such a  unified cosmological model, that is  quantified by only a single parameter $\mu$, that can be considered as a minimal extension of the $\Lambda$-cold dark matter cosmology. We investigate its observational boundaries together with an analysis of the universe at large scale. Our study shows that at early time the model behaves like a dust, and as time evolves, it mimics a dark energy fluid depicting a clear transition from the early decelerating phase to the late cosmic accelerating phase. Finally, the model approaches  the cosmological constant boundary in an asymptotic manner. 
We remark that for the present unified model, the estimations of $H_0$ are slightly higher than its local estimation and thus alleviating the $H_0$ tension.  
\end{abstract}

\maketitle
\section{Introduction}

Dark energy and dark matter are supposedly two most important constituents  of our universe comprising  about 96\% of the total energy density of our universe \cite{Ade:2015xua} while the known matter makes up the remaining $\sim $ 4\%. The origin and  the nature of these dark fluids are not so well-known despite of many observational missions performed by several space projects. Sometimes, it is believed that both dark matter and dark energy evolve separately, that means, the dynamics of each dark fluid is independent of the other. On the other hand, it is also conjectured that dark matter and dark energy are actually coming from a single entity that could play the role of both dark fluids. 
The single dark fluid exhibiting two different dark sides of the universe is commonly known as the  {\it unified dark fluid}.  The unified dark fluids are very well known  and got massive attention to the cosmological community. 
The Chaplygin gas model \cite{Chaplygin,Kamenshchik:2001cp, Bilic:2001cg,Gorini:2002kf,gorini01} and its modified versions, namely the generalized Chaplygin model \cite{Bento:2002ps,Amendola:2003bz,Avelino:2003cf,Multamaki:2003ed,Lu:2009zzf,Xu:2012qx}  and the modified Chaplygin model \cite{Benaoum:2002zs,Debnath:2004cd,Lu:2008zzb,Xu:2012ca}, and several other extensions of Chaplygin cosmology \cite{Guo:2005qy} were successively introduced in the literature and consequently they had been investigated by several investigators aiming to test their observational viabilities. Moreover, the Chaplygin type fluids have also been found to explain the inflationary expansion of the universe while Chaplygin cosmologies can follow from scalar field cosmologies for specific scalar field potentials, see \cite{BouhmadiLopez:2011kw,delCampo:2008vr,Herrera:2016sov,Barrow:2016qkh,an01,an02,an03}. We also refer to some earlier works providing with a Lagrangian description to the Chaplygin cosmologies \cite{Banerjee:2005vy,Banerjee:2006na}. We also refer to some recent works in the context of unified cosmological models, see for 
instance \cite{Basilakos:2008jb, Lima:2012mu, Perico:2013mna, Basilakos:2013xpa, Kofinas:2017gfv,Anagnostopoulos:2018jdq}.

The Chaplygin gas has the equation of state $p =- A/\rho$ ($A > 0$), where $p$ and $\rho$ are respectively the pressure and the energy density of the fluid. 
Subsequently, this simplest unified cosmological model was modified leading to generalized Chaplygin gas ($p = -A/\rho^{\alpha}$, $A > 0$ and $\alpha \geq 0$), modified Chaplygin gas ($p = A \rho -B/\rho^{\alpha}$ where $A$, $B$, $\alpha$ are any free parameters).  In all three Chaplygin gas models,  a common property that we observe is that,  if one works in a homogeneous and isotropic background of the universe which is supported by the observational evidences (described by the Friedmann-Lema\^{i}tre-Robertson-Walker line element), 
then at early time this model behaves like a dust fluid and in the late time of the universe, a cosmological constant type fluid it appears. Between the three different models, the modified Chaplygin model can also behave like a radiation fluid for $A = 1/3$, $B =0$; such a property is absent in other two Chaplygin models. With the developments in the astronomical data, the Chaplygin-gas models have been examined in detail in the literature.  However, apart from the Chaplygin gas models, one may also consider some alternative unified dark fluid models that could exhibit similar qualities to that of the Chaplygin type models. 

Thus, following this motivation, in the present work we investigate a different unified dark model (we call it as a unified model, `UM' in short) in order to mainly examine the large scale structure formation of the universe in the context of the new unified cosmological fluid as well as to compare its potentiality with resoect to the well known Chaplygin models. In particular, we consider a spatially flat Friedmann-Lema\^{i}tre-Robertson-Walker (FLRW) metric as the underlying geometry where the matter sector is minimally coupled to the Einstein gravity. 

The work is organized in the following way. In section \ref{sec-field} we describe the gravitational field equations at the background and perturbation levels for a unified cosmological model. In section \ref{sec-data} we describe the observational data that we have used to constrain the model and the results of the analyses where we perform several observational datasets. Finally, in section \ref{sec-conclu} we conclude our work with a brief summary.

\section{Field equations}
\label{sec-field}

In the large scale, our universe is homogeneous and isotropic and its geometry is best described by the Friedmann-Lema\^{i}tre-Robertson-Walker (FLRW) metric 
\begin{eqnarray}
ds^2 = -dt^2 + a^2 (t) \left[\frac{dr^2}{1-Kr^2} + r^2 \left(d\theta^2 + \sin^2 \theta d \phi^2\right) \right], 
\end{eqnarray}
where $a(t)$ is the expansion scale factor of the universe and $K \in {0, +1, -1}$ is the curvature scalar. For $K =0$, we have a spatially flat universe, for $K=1$, the universe is closed and for $K = -1$, it is open. 

We assume that the gravity sector of the universe is described by the Einstein gravity and the matter sector is minimally coupled to gravity where no such interaction exists between any two matter fluids. In particular, we consider that the total energy density of the universe is shared by three different fluids, namely, radiation, baryons and a unified fluid that acts the role of both dark energy at late time and dark matter at early time. Thus, the total energy density is defined by, $\rho_{\rm tot} = \rho_r+\rho_b +\rho_u$, where  $\rho_r$, $\rho_b$ and $\rho_u$ are the energy density of radiation, baryons and a unified cosmic fluid respectively. Similarly, by $p_{\rm tot} = p_r +p_b +p_u$, we describe the total pressure of the fluids where $p_r$, $p_b$ and $p_u$ respectively denote the pressure of radiation, baryons and the unified dark fluid. Now,  in such a spacetime, one can write down  Einstein's field equations  as 

\begin{eqnarray}
H^2  + \frac{K}{a^2} = \frac{8 \pi G}{3} \rho_{\rm tot},\label{efe1}\\
2 \dot{H} + 3 H^2 + \frac{K}{a^2} = - 8 \pi G p_{\rm tot},\label{efe2}
\end{eqnarray}
which are two independent field equations; here an overhead dot represents the cosmic time differentiation and  $H \equiv \dot{a}/a$ is the Hubble rate of the FLRW universe.  Because the observational data always favor a spatially flat universe, thus, throughout the work we shall assume $K =0$. 
As there is no interaction between any two fluids, thus,  the conservation equation for the $i$-th fluid (where $i = r, b, u$) follows, $\dot{\rho}_i + 3 H (p_i+\rho_i) = 0$. 
The equation-of-state of radiation is $p_r = \rho_r/3$ and for baryons it is $p_b  =0$. Thus, once the equation-of-state of the unified fluid is prescribed, then using conservation equation of each fluid together with the Hubble equation (\ref{efe1}), the dynamics of the universe can in principle be determined. Hence, determining an equation-of-state for the unified fluid is a challenging issue. Interestingly, it has been already found that the unified fluids, such as Chaplygin and its successive generalizations can be derived from a field theoretic description. It is easy to show that from an action formalism representing a tachyon field as \cite{Hova:2010na}:

\begin{eqnarray}
\mathcal{S}  = - \int d^4 x V (\phi) \sqrt{\det (g_{\mu \nu} + \partial_{\mu} \phi \partial_{\nu} \phi)}~~, 
\end{eqnarray}
where $V (\phi)$ is the potential of the tachyon field, in a flat FLRW universe, the energy density ($\rho_{\phi}$) and pressure ($p_{\phi}$) of that field can be derived to be 
\begin{eqnarray}
\rho_{\phi} = \frac{V(\phi)}{\sqrt{1- \dot{\phi}^2}},\qquad p_{\phi} = - V (\phi) \sqrt{1- \dot{\phi}^2}, 
\end{eqnarray}
from which one can find the equation of state of the tachyon field,  $p_{\phi} = - V^2(\phi)/\rho_{\phi}$. This equation mimicks the well known equation-of-state for the Chaplygin gas if the potential is assumed to be constant. In the present work we shall introduce a different equation-of-state  of a unified dark fluid proposed in \cite{Hova:2010na} and this unified model has a field theoretic origin. In particular, it has been shown that the model has an equivalent tachyonic potential \cite{Hova:2010na}. 
The explicit expression of the equation-of-state of the unified dark fluid is  \cite{Hova:2010na,Hernandez-Almada:2018osh}:

\begin{eqnarray}\label{eos1}
p_u = -\rho_u +  \rho_u \, {\rm sinc}\left(\frac{\mu \pi \rho_{u,0}}{\rho_u} \right) 
\end{eqnarray}
where ${\rm sinc}(\theta) = \sin \theta /\theta$; $\mu \neq 0$ is any dimensionless quantity, $\rho_{u,0}$ is the present value of the unified dark fluid. 
One can quickly realize that the proposed equation of state (\ref{eos1}) has a very general structure having the following equation of state, $p_{u} = - \rho_{u} + f(\rho_{u})$, where $f(\rho_u)$ is any analytical function of $\rho_u$ and for $f(\rho_{u}) = 0$, the cosmological constant scenario (i.e., $p_u = - \rho_u$) is recovered. The choice $f(\rho_u) =  {\rm sinc}\left(\mu \pi \rho_{u,0}/\rho_u  \right)$ has some interesting features. Considering its first order expansion, from (\ref{eos1}), one finds that, $p_u \sim 0$, that means the unified model catches the matter dominated era. While on the other hand, if we consider up to the second order expansion of $f(\rho_u) =  {\rm sinc}\left(\mu \pi \rho_{u,0}/\rho_u  \right)$, then eqn. (\ref{eos1}) reduces into $p_u  = - A/ \rho_{u}^3$, where $A = \frac{\mu \pi \rho_{u,0}^3}{3!}$. This represents an exotic fluid resembling with Chaplygin gas model.  
At this point consider the field equations (\ref{efe1}), (\ref{efe2}) where only the unified fluid exists. Let $H\left( t\right) = H_e t^{-1}$ ($H_e$ is a constant) describes an ideal gas solution. Recall that for $H_e =\frac{2}{3}$, the ideal gas is a dust fluid. From (\ref{efe1}) and for a spatially flat FLRW universe, if follows that $\rho _{u}=H_e^{2}t^{-2}$. Hence,  equation (\ref{efe2}) becomes%
$$
-\frac{H_e}{t^{2}}\left( 2\left( H_e -1\right) + H_e^{3}\frac{\sin
\left( \frac{\mu \pi \rho _{u,0}}{H_e^{2}}t^{2}\right) }{t^{2}}\right) =0,
$$
hence the later equation has a solution at the early universe when $%
t\rightarrow 0$ if and only if $\mu \pi \rho _{u,0}=\frac{2\left(
H_e-1\right) }{H_e}$, where for the case of a dust fluid we find $\mu
\pi \rho _{u,0}=1$.

In order to study the evolution of the ideal gas solution at early times we
substitute $H\left( t\right) = H_e t^{-1}+\varepsilon ~t H_{p}\left( t\right)
$ in (\ref{efe1}), (\ref{efe2}) and around the value $\varepsilon
\rightarrow 0$ near to $t\rightarrow 0$ we find
$$
tH_{p}^{\prime }+6H_{p}=0~,
$$
that is
$$
H\left( t\right) = H_e t^{-1}+\varepsilon t^{-5}~,
$$
from where we can infer that the perturbations decay at early times. The latter solution holds only for small values of $t$ hence we can not infer about the stability of the solution.

Now, using the conservation equation, one can solve the evolution of the unified dark fluid as follows
\begin{eqnarray}\label{energy-density}
\rho_u=\rho_{u0} \left( \frac{\mu\pi}{2 \arctan \left[a^3 \tan(\mu\pi/2)\right]} \right),
\end{eqnarray}
and consequently, the explicit expression for the equation of state $w_u = p_u/\rho_u$ of this dark fluid, i.e., (\ref{eos1}) becomes, 
\begin{eqnarray}\label{eos-explicit}
w_u=-1+\frac{a^{-3}\tan(\mu\pi/2)}{[a^{-6}+\tan^2(\mu\pi/2)]\arctan[a^3\tan(\mu\pi/2)]}
\end{eqnarray}

We now proceed towards the graphical presentation of the cosmological parameters associated with this unified model and also we perform a comparison of the present model with the well known Generalized Chaplygin Gas (GCG) model. The common feature of the present unified model with the GCG model is that both of them are quantified by a single parameter. This kind of comparison is useful to understand how the present model is qualitatively close to the already known unified models for years. 

\begin{figure*}
\includegraphics[width=0.4\textwidth]{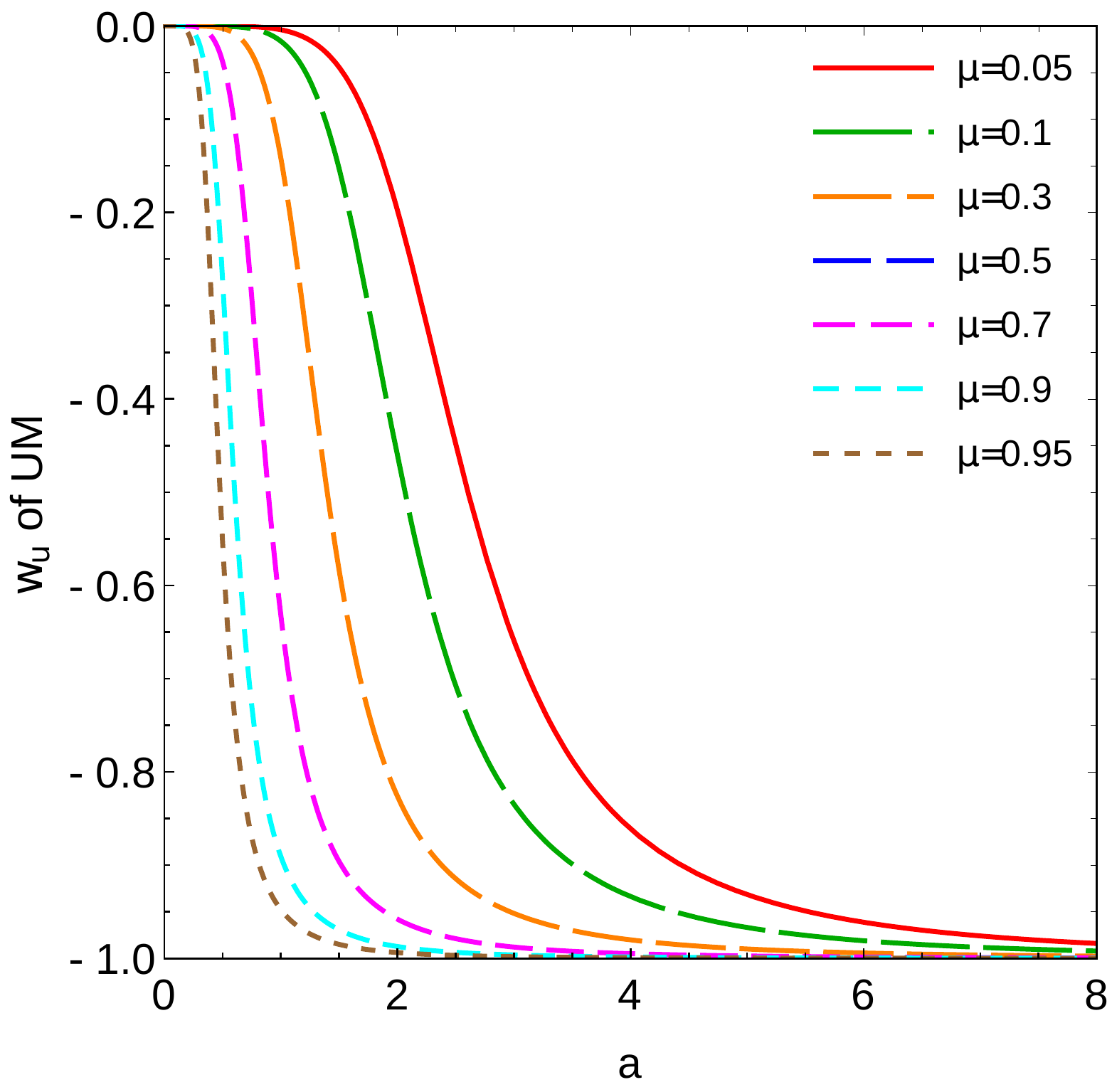}
\includegraphics[width=0.4\textwidth]{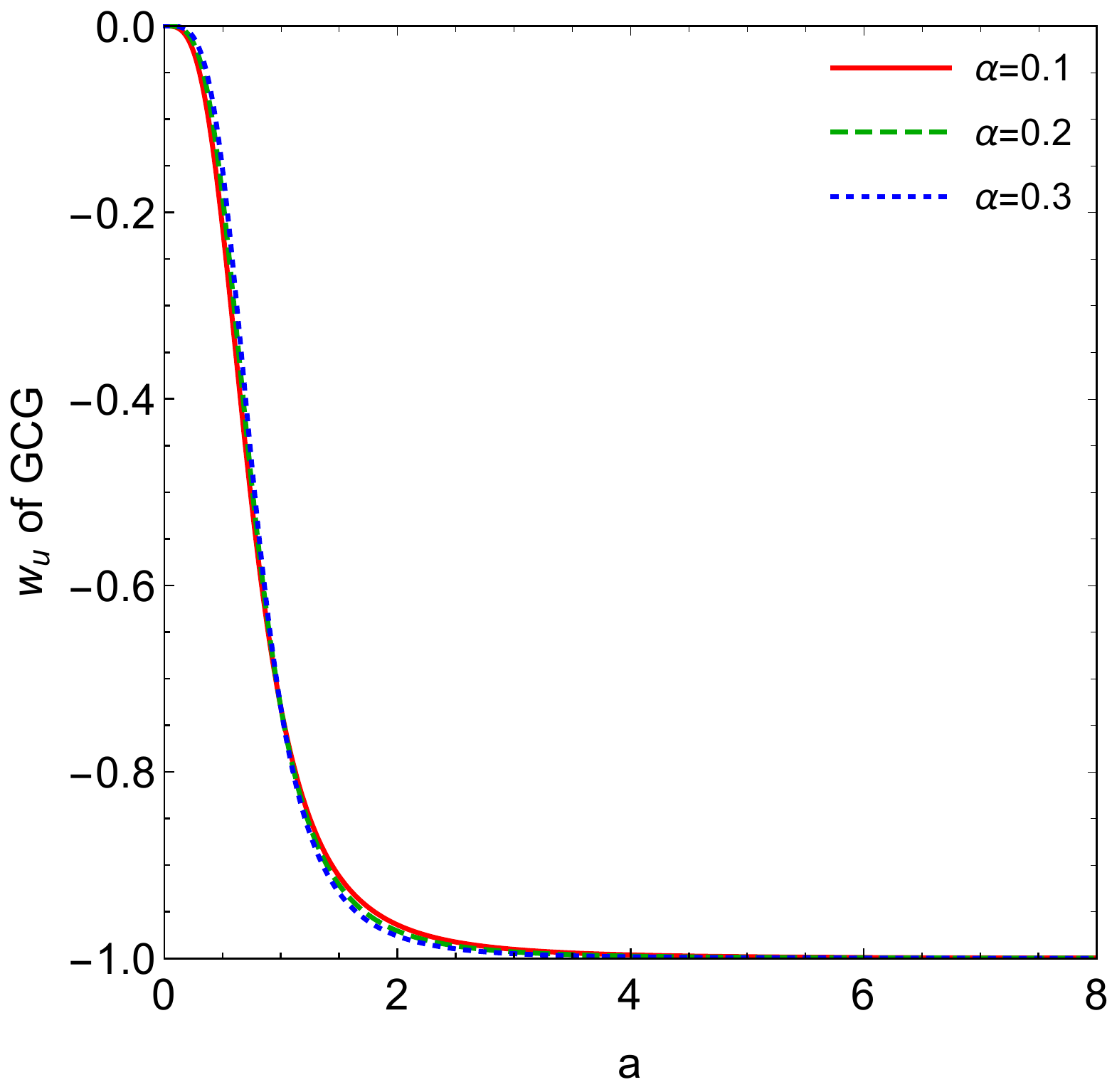}
\caption{The evolution of the equation of state for the unified cosmic fluid (\ref{eos-explicit}) [left graph] and of the GCG model [right graph] have been shown using different values of the key parameters, namely, $\mu$ of the present model and $\alpha$ of the GCG model.  From both the graphs, one can see that the evolution of these fluids seem to be similar in which the equation of state has a smooth transition 
from  the dust fluid to the cosmological constant dominated universe ($w =-1$). }
\label{fig-eos}
\end{figure*}

In Fig. \ref{fig-eos} we present the equation of state for the present unified model and the generalized Chaplygin model. The left panel of Fig. \ref{fig-eos} stands for the present UM while the right panel stands for GCG. In both the panels we have used various values of the corresponding key parameters, namely $\mu$ and $\alpha$. From these plots, one can notice their evolution are slightly different but qualitatively they are same, in the sense that both of them correctly prescribe the dust era in the early phase and finally a cosmological constant dominated are asymptotically recovered.   

\begin{figure*}
\includegraphics[width=0.45\textwidth]{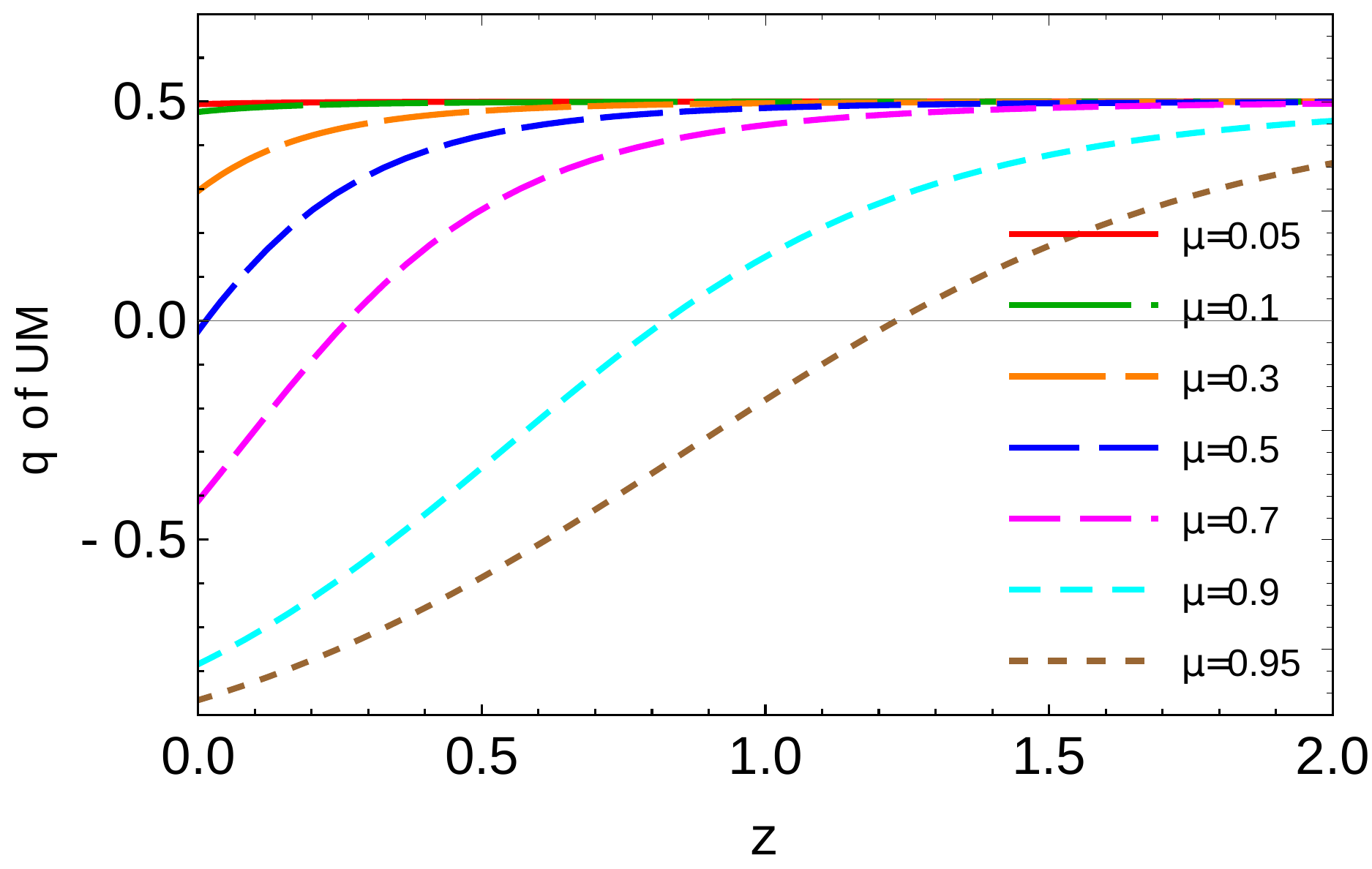}
\includegraphics[width=0.45\textwidth]{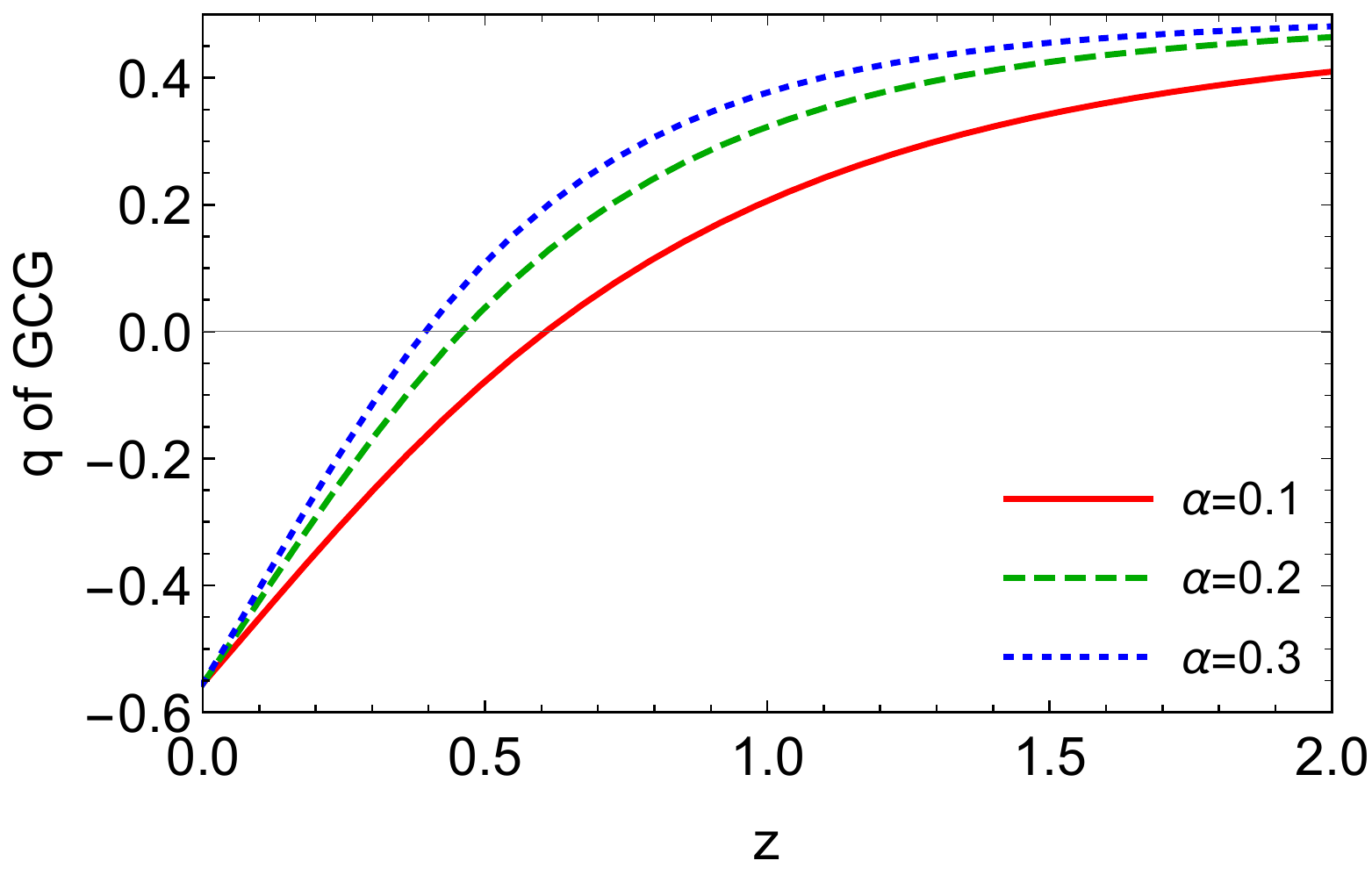}
\caption{The evolution of the deceleration parameter for the unified cosmic fluid (\ref{eos-explicit}) [left graph] and of the GCG model [right graph] have been shown using different values of the key parameters, namely, $\mu$ of the present model and $\alpha$ of the GCG model. }
\label{fig-deceleration}
\end{figure*}

In Fig.  \ref{fig-deceleration} we depict the evolution of the deceleration parameter for both the unified models (i.e., the present unified model and the known GCG model) which is quite interesting in the sense that the free parameter $\mu$ has an effective role in describing the transition from the past decelerating phase to the present one.
From both the plots of Fig. \ref{fig-deceleration}, we can find a smooth transition from the past decelerating era to the current accelerating phase, however, one can also notice from the left plot of Fig. \ref{fig-deceleration} that all values of $\mu$ cannot predict the transition from $q> 0$ to $q< 0$ phase. In fact, we see that for $\mu \geq 0.7$, we have a correct picture of the present universe. This actually gives a restriction on $\mu$.

\begin{figure*}
\includegraphics[width=0.45\textwidth]{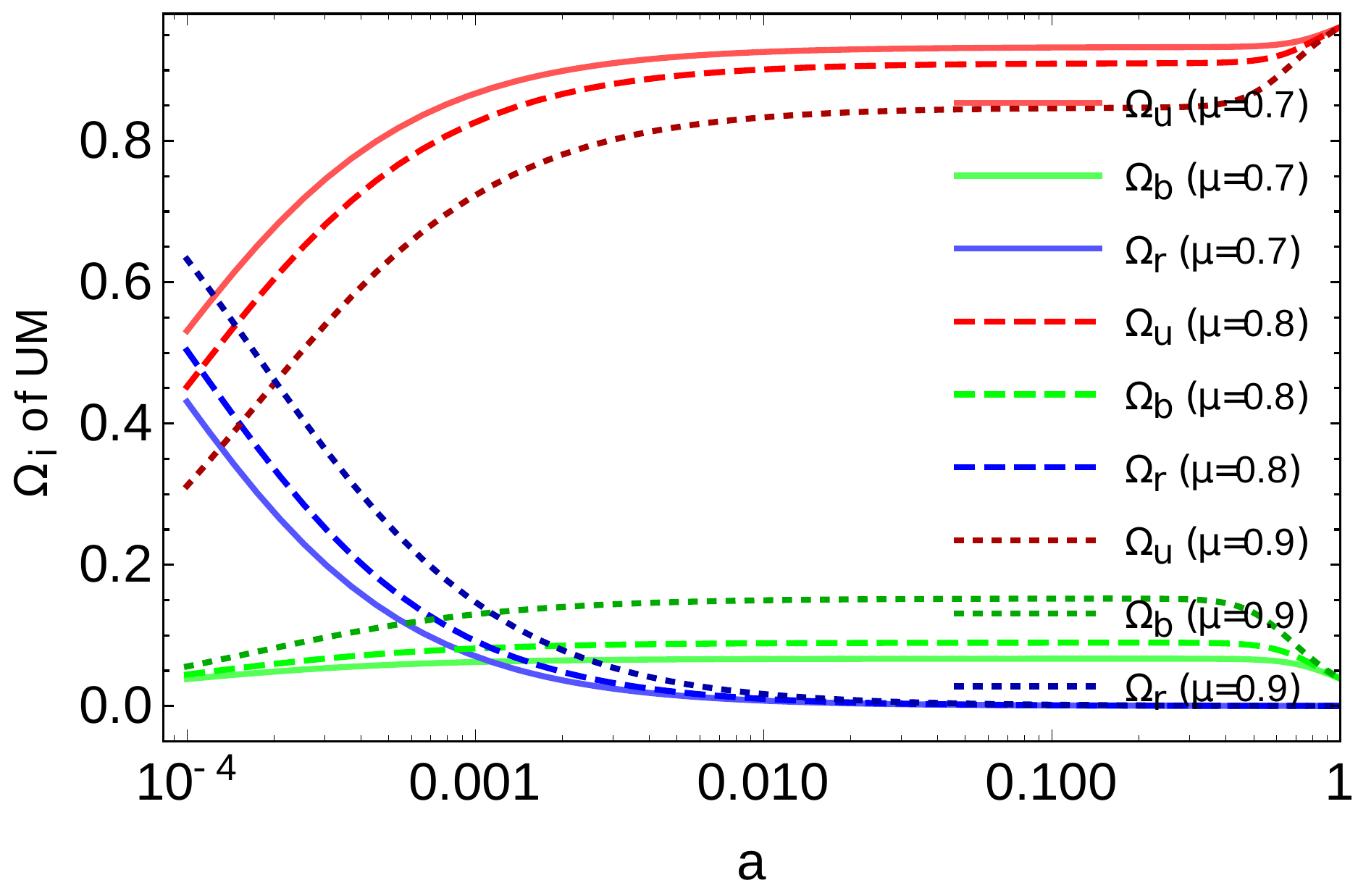}
\includegraphics[width=0.45\textwidth]{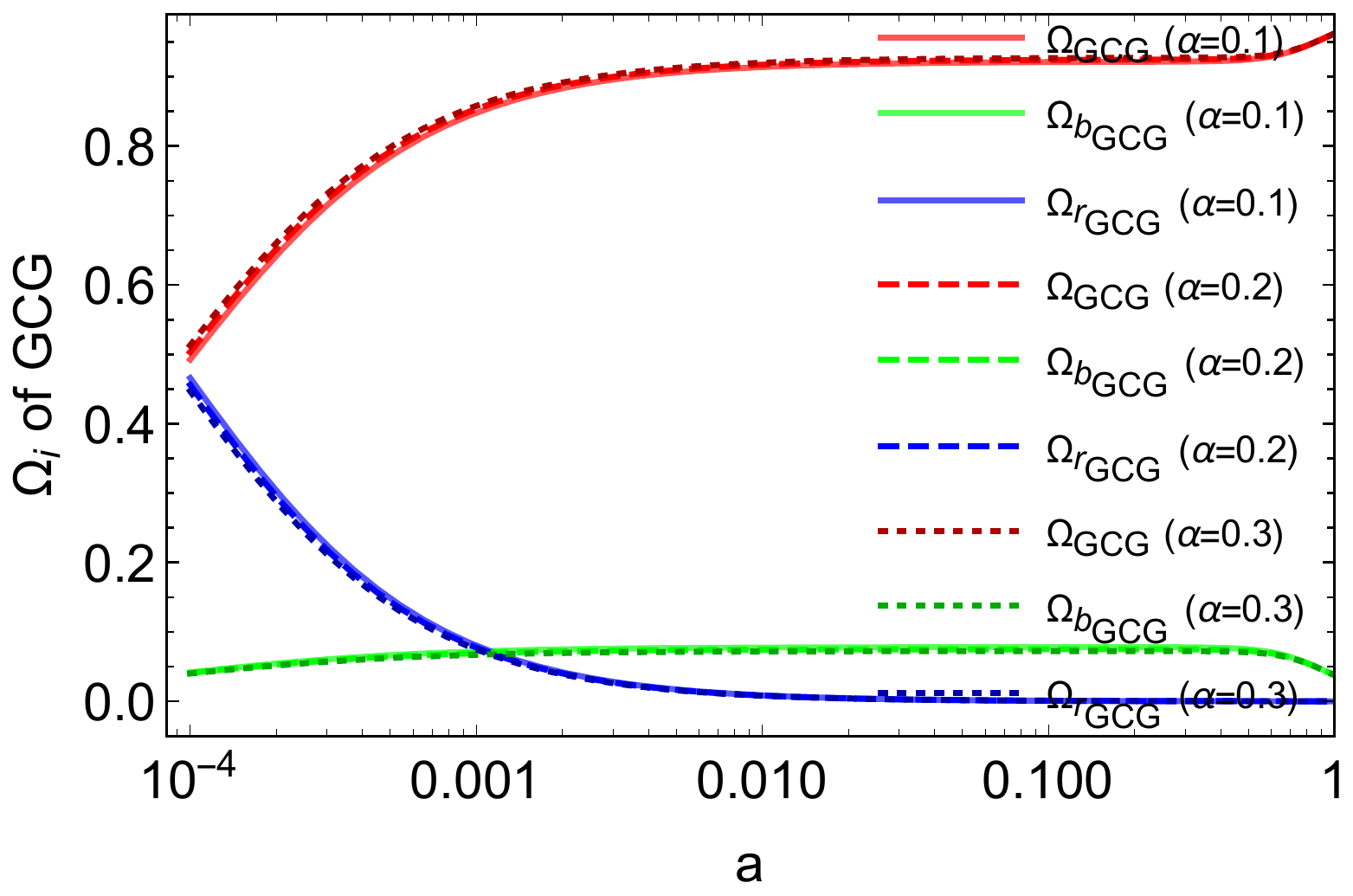}
\caption{The evolution of the density parameters for the unified cosmic scenario (\ref{eos-explicit}) [left graph] and of the GCG scenario [right graph] have been shown using different values of the key parameters, namely, $\mu$ of the present model and $\alpha$ of the GCG model. }
\label{fig-Omega}
\end{figure*}

Finally, in Fig. \ref{fig-Omega} we have shown the dimensionless density parameters $\Omega_u$, $\Omega_b$ and $\Omega_r$ for both the unified scenarios. The left panel of Fig. \ref{fig-Omega} corresponds to present UM while the right panel of Fig. \ref{fig-Omega} corresponds to GCG.


Having presented the background evolution of the unified cosmic model, we proceed towards its perturbation analysis. We consider the perturbed metric in the synchronous gauge that takes the form 
\cite{Mukhanov,Ma:1995ey,Malik:2008im}

\begin{eqnarray}
ds^2 = a^2 (\tau) \left[- d \tau^2 + \left(\delta_{ij} + h_{ij}\right) dx^i d^j\right]
\end{eqnarray}
where $\tau$ is the conformal time, and $h_{ij}$ is the metric perturbations while $\delta_{ij}$ is the unperturbed part of the metric tensor.  We note that there is absolutely no objection to consider the perturbations equations in the conformal Newtonian gauge. However, most of the calculations of linear growth fluctuations have been carried out in the synchronous gauge and they are easily included in the Code for Anisotropies in the Microwave Background (CAMB). Thus, to work with synchronous gause, it is easy to modify the original CAMB. While the advantage of Newtonian gauge \footnote{The perturbations in the confomal Newtonian gauge are characterized by two scalar potentials $\Psi$ and $\Phi$ with the line element \cite{Ma:1995ey}: $ds^2  = a^2 (\tau) \left[-(1+2 \Psi) d\tau^2  + (1 - 2 \Phi) dx^{i} dx_{i} \right]$, where $\tau$ is the conformal time. For more details, please see \cite{Ma:1995ey}. } is that the metric tensor $g_{\mu\nu}$ becomes diagonal. This simplifies the calculations and leads to simple geodesic eqautions. Another advantage of the Newtonian gauge is that the scalar potential, $\Psi$, plays the role of the gravitational potential in the Newtonian limit, and thus, has a simple physical interpretation. Although we note that these two gauges could be transformed from one another by the gauge transforamtion \cite{Ma:1995ey}. We also refer to \cite{Basilakos:2012ut} for more discussions in this regard. However, in the present work we shall work in the synchronous gauge. 
Now, using the conservation equations $T^{\mu \nu}_{; \nu} = 0$, one can now find the gravitational field equations
using this perturbed metric. For the unified dark fluid, one can write down the density and velocity perturbations for a mode with 
wavenumber ${k}$ as \cite{Xu:2012zm}: 

\begin{eqnarray}
\delta_{u}^{\prime} = -(1+w_u) \left(\theta_u + \frac{h}{2}\right)- 3 \mathcal{H} \left(\frac{\delta p_u}{\delta \rho_{u}} - w_u\right) \delta_u,\\
\theta_u^{\prime} = - \mathcal{H}  \left(1- 3 c^2_{s(u),ad}\right) + \left(\frac{\delta p_u/\delta \rho_{u}}{1+w_u}\right) k^2 \delta_u - k^2 \sigma_u,
\end{eqnarray}
where the primes denote the derivatives with respect to the conformal time $\tau$; $\mathcal{H}$ is the conformal Hubble factor;  $\delta_u = \delta \rho_{u}/\rho_{u}$ is the density perturbations; $\theta_{u}$ is the divergence of the unified fluid 
velocity; $h = h^{j}_{j}$ is the trace of the metric perturbations $h_{ij}$; $c^2_{s(u),ad}$  is the adiabatic sound speed of the unified fluid taking the expression $c^2_{s(u),ad}  = p_u^{\prime}/\rho_{u}^{\prime} =  w_u - \frac{w_u^{\prime}}{3 \mathcal{H}(1+w_u)}$; and $\sigma_u$ is the shear perturbations for the unified fluid. One can notice that the adiabatic sound speed for the unified fluid, namely, $c^2_{s(u),ad}$ could be negative (that means $c_{s(u),ad}$ becomes an imaginary number)
and as a result we will have instabilities in the perturbations. For instance, $w_u = $ constant, $c^2_{s(u),ad} <0$ and similarly for other equations of state, this could equally happen. Thus, we need to find an alternative way out in order to bypass such instabilities in the perturbations. A possible way that removes such problem is to allow an entropy perturbation into the framework and assume either positive or null effective sound speed (sum of the adiabatic and entropic one).  We thus follow the formalism \cite{Hu:1998kj} developed for a
generalized dark matter. Using this formalism,  the entropy perturbation for the unified fluid can be separated out where $p_u \Gamma_u  = \delta p_u - c^2_{s(u),ad} \delta \rho_d$, which is gauge independent and $\Gamma_u$ is a constant. Now in the rest frame of the unified fluid, the entropy perturbations is characterized by $w_u \Gamma_u  = \left(c^2_{s(u),eff} - c^2_{s(u),ad}\right) \delta_u^{\rm rest}$, where $c^2_{s(u),eff}$ is the effective sound speed of the unified fluid and $\delta_u^{\rm rest}$ in terms of an arbitrary gauge is specified as $\delta_u^{\rm rest} = \delta_u + 3 \mathcal{H} (1+w_u) \frac{\theta_{u}}{k^2}$ and this is a gauge-invariant form for the entropy perturbations. Now, with this set up the density and velocity perturbations for the unified dark fluid now  takes the form 

\begin{eqnarray}
\dot{\delta}_u=-(1+w_u)(\theta_u+\frac{\dot{h}}{2})-3\mathcal{H}(c^2_{s,eff}-w_u)\delta_u \nonumber\\-9\mathcal {H}^2(c^2_{s,eff}-c^2_{s,ad})(1+w_u)\frac{\theta_u}{k^2}, \\
\dot{\theta}_u=-\mathcal{H}(1-3c^2_{s,eff})\theta_u+\frac{c^2_{s,eff}}{1+w_u}k^2\delta_u-k^2\sigma_u~.
\end{eqnarray}
During the analysis we have neglected the shear perturbations for the unified fluid in agreement with the earlier works \cite{Xu:2012zm}, that means we set $\sigma_u = 0$.  Usually, one can consider the nonzero $\sigma_u$ for a more generalized version, however, its inclusion extends the parameters space and the degeneracy between other parameters increases. Thus, we exclude its possibility from this picture.  
Additionally, we assume the effective sound speed to be null.

\begin{table}
\begin{center}
\begin{tabular}{c|c|}
Parameter                    & Prior\\
\hline 
$\Omega_{b} h^2$             & $[0.005,0.1]$\\
$\tau$                       & $[0.01,0.8]$\\
$n_s$                        & $[0.5, 1.5]$\\
$\log[10^{10}A_{s}]$         & $[2.4,4]$\\
$100\theta_{MC}$             & $[0.5,10]$\\ 
$\mu$                        & $[0.01, 2]$\\ 
\end{tabular}
\end{center}
\caption{Flat priors on various free parameters of the unified model have been shown. }
\label{tab:priors}
\end{table}

\begingroup                                                                                                                     
\begin{center}                                                                                                                  
\begin{table*}                                                                                                                   
\begin{tabular}{cccccc}                                                                                                            
\hline\hline                                                                                                                    
Parameters & CC & Pantheon & Pantheon+CC \\ \hline

$\Omega_b h^2$ & $    0.01915_{-    0.01415-    0.01415}^{+    0.00376+    0.00839}$ & $    0.04201_{-    0.02989-    0.03701}^{+    0.01934+    0.03536}$ & $    0.02336_{-    0.00335-    0.00805}^{+    0.00430+    0.00815}$\\

$\mu$ & $    0.849_{-    0.019-    0.062}^{+    0.033+    0.053}$ & $    0.825_{-    0.015-    0.033}^{+    0.017+    0.033}$ & $    0.826_{-    0.012-    0.025}^{+    0.013+    0.023}$ \\

$H_0$ & $   70.23_{-    2.75-    6.80}^{+    3.43+    5.97}$ & $   78.12_{-   16.24-   20.06}^{+   10.19+   22.84}$ & $   68.18_{-    1.82-    3.75}^{+    1.85+    3.70}$  \\
\hline\hline

$\chi^2_{\rm best-fit}$ &  14.542 & 1036.41 & 1052.144 \\
\hline 
\hline                                                                                                                    
\end{tabular}                                                                                                                   
\caption{68\% and 95\% confidence-level constraints on the cosmological parameters of the present unified model using CC, Pantheon and Pantheon+CC datasets. }
\label{tab:background}                                                                                                   
\end{table*}                                                                                                                     
\end{center}                                                                                                                    
\endgroup  

\begin{figure*}
\includegraphics[width=0.25\textwidth]{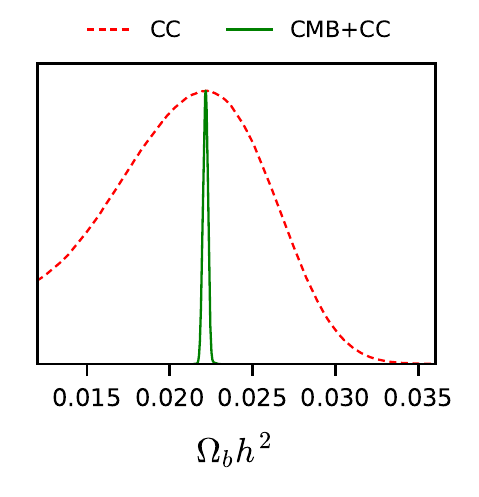}
\includegraphics[width=0.25\textwidth]{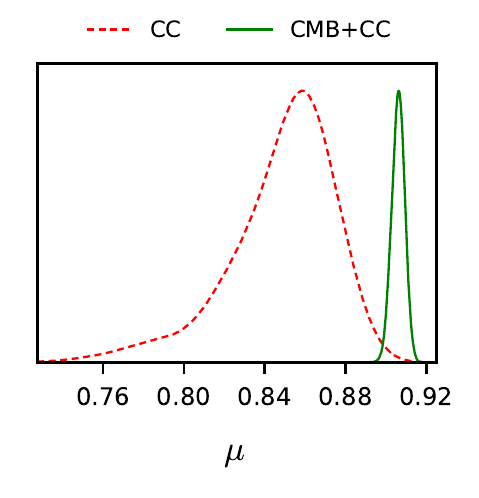}
\includegraphics[width=0.25\textwidth]{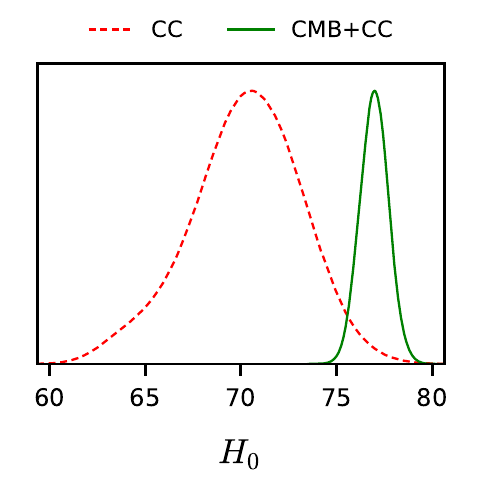}
\includegraphics[width=0.25\textwidth]{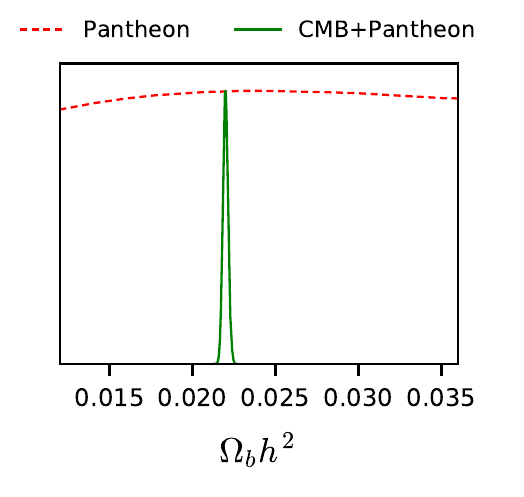}
\includegraphics[width=0.25\textwidth]{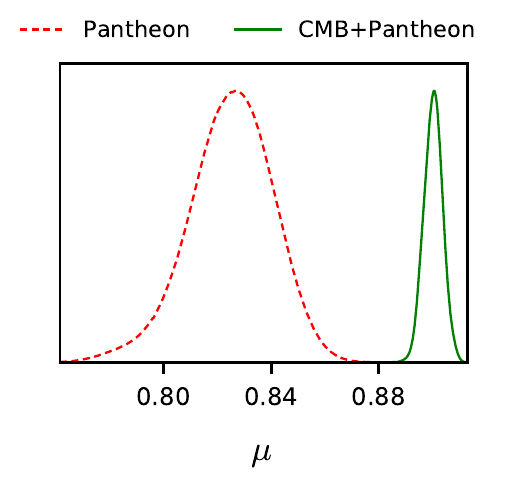}
\includegraphics[width=0.25\textwidth]{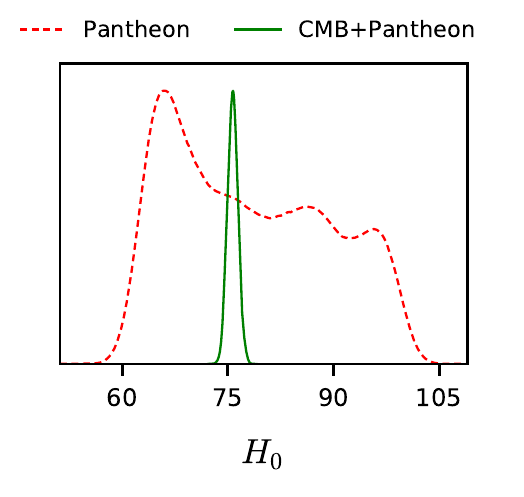}
\includegraphics[width=0.25\textwidth]{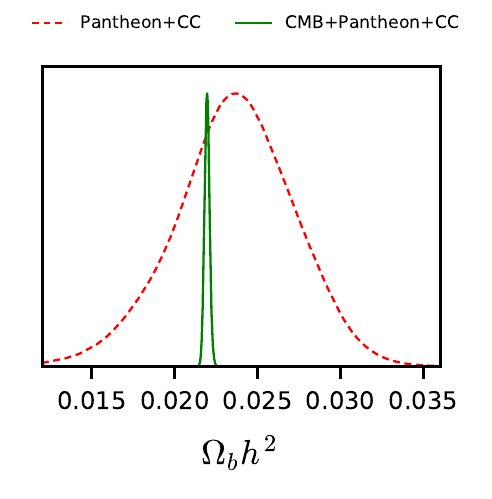}
\includegraphics[width=0.25\textwidth]{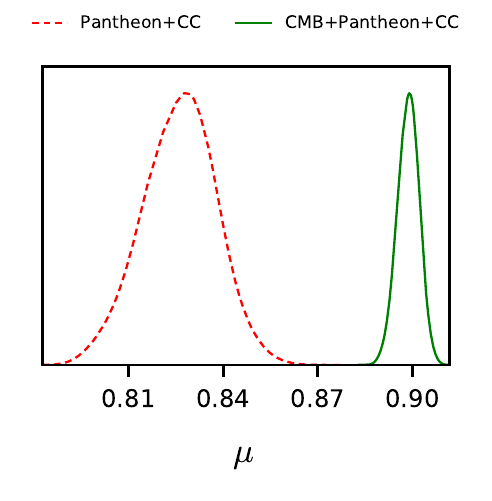}
\includegraphics[width=0.25\textwidth]{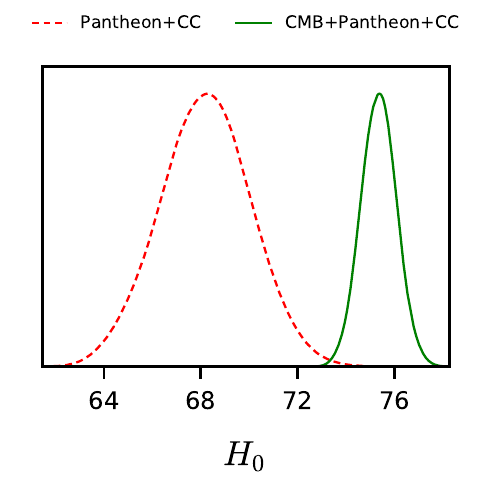}
\caption{We compare the observational constraints through the one-dimensional posterior distrbibutions of some parameters of the unified model before and after the inclusion of CMB data with the background datasets, namely CC, Pantheon and Pantheon+CC. }
\label{fig:background}
\end{figure*}

\begingroup                                                                                                                     
\begin{center}                                                                                                                  
\begin{table*}                                                                                                                   
\begin{tabular}{ccccccccccccc}                                                                                                            
\hline\hline                                                                                                                    
Parameters & CMB &  CMB+CC & CMB+Pantheon & CMB+Pantheon+CC \\ \hline
$\Omega_b h^2$ & $    0.02220_{-    0.00015-    0.00030}^{+    0.00015+    0.00030}$ & $    0.02215_{-    0.00015-    0.00030}^{+    0.00015+    0.00028}$ &  $    0.02198_{-    0.00016-    0.00032}^{+    0.00015+    0.00032}$ & $    0.02194_{-    0.00015-    0.00029}^{+    0.00014+    0.00031}$  \\

$100\theta_{MC}$ & $    1.02351_{-    0.00034-    0.00067}^{+    0.00035+    0.00067}$ & $    1.02361_{-    0.00035-    0.00070}^{+    0.00036+    0.00067}$  & $    1.02395_{-    0.00034-    0.00064}^{+    0.00034+    0.00065}$ & $    1.02405_{-    0.00036-    0.00068}^{+    0.00032+    0.00070}$ \\

$\tau$ & $    0.075_{-    0.016-    0.033}^{+    0.016+    0.033}$ & $    0.069_{-    0.017-    0.036}^{+    0.019+    0.033}$ & $    0.060_{-    0.016-    0.031}^{+    0.016+    0.032}$ & $    0.057_{-    0.016-    0.032}^{+    0.016+    0.032}$ \\

$n_s$ & $    0.9658_{-    0.0044-    0.0087}^{+    0.0044+    0.0089}$ & $    0.9638_{-    0.0048-    0.0087}^{+    0.0043+    0.0090}$ & $    0.9578_{-    0.0049-    0.0086}^{+    0.0043+    0.0090}$ & $    0.9564_{-    0.0042-    0.0088}^{+    0.0043+    0.0086}$ \\

${\rm{ln}}(10^{10} A_s)$ & $    3.085_{-    0.032-    0.065}^{+    0.033+    0.064}$ & $    3.076_{-    0.033-    0.070}^{+    0.038+    0.069}$ & $    3.062_{-    0.031-    0.062}^{+    0.030+    0.062}$ & $    3.058_{-    0.032-    0.061}^{+    0.032+    0.062}$ \\

$\mu$ & $    0.908_{-    0.0029-    0.0064}^{+    0.0033+    0.0058}$ & $    0.906_{-    0.0031-    0.0064}^{+    0.0035+    0.0062}$ & $    0.900_{-    0.0035-    0.0068}^{+    0.0034+    0.0069}$ & $    0.899_{-    0.0035-    0.0071}^{+    0.0037+    0.0067}$  \\

$H_0$ & $   77.33_{-    0.73-    1.48}^{+    0.71+    1.41}$ & $   76.97_{-    0.74-    1.44}^{+    0.79+    1.47}$ & $   75.68_{-    0.76-    1.49}^{+    0.74+    1.52}$  & $   75.35_{-    0.73-    1.45}^{+    0.71+    1.47}$ \\
\hline
$\chi^2_{\rm best-fit}$ & 12965.656 & 12987.686 & 14052.544 & 14071.64 \\
\hline\hline                                                                                                                    
\end{tabular}                                                                                                                   
\caption{68\% and 95\% confidence-level constraints on the parameters of the unified cosmological model have been shown for different observational datasets. Here $H_0$ is in the units of km/s/Mpc.  }
\label{tab:results1}                                                                                                   
\end{table*}                                                                                                                     
\end{center}                                                                                                                    
\endgroup    
\begin{figure*}
\includegraphics[width=0.7\textwidth]{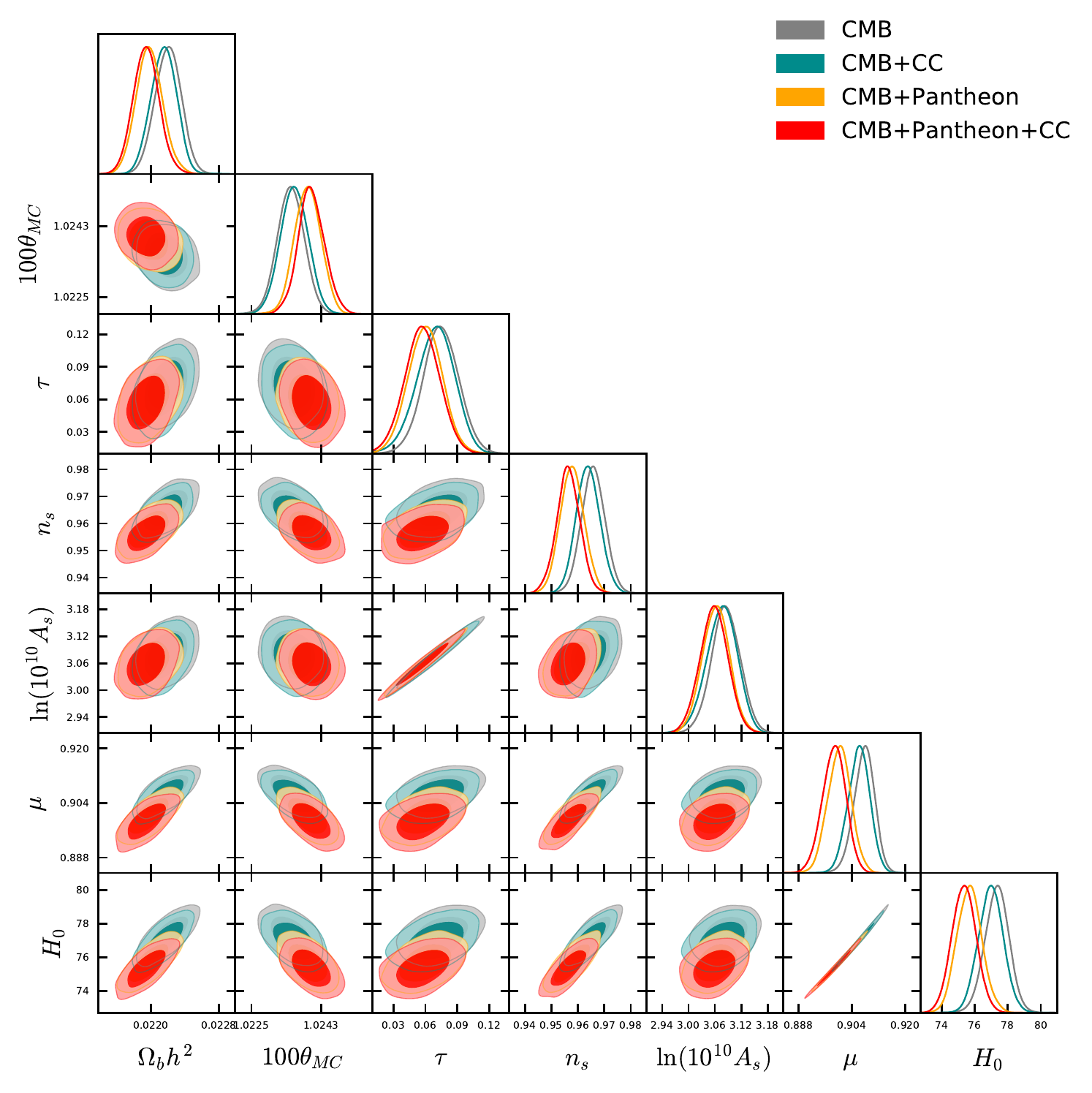}
\caption{68\% and 95\% confidence-level contour plots between the free parameters of the unified model as well as the one dimensional marginalized posterior distributions for all the model parameters using several combinations of the datasets. From this figure one can clearly see that $\mu$ has a very strong positive correlation with $H_0$ irrespective of the observational datasets. }
\label{fig-2D}
\end{figure*}
\begin{figure*}
\includegraphics[width=0.35\textwidth]{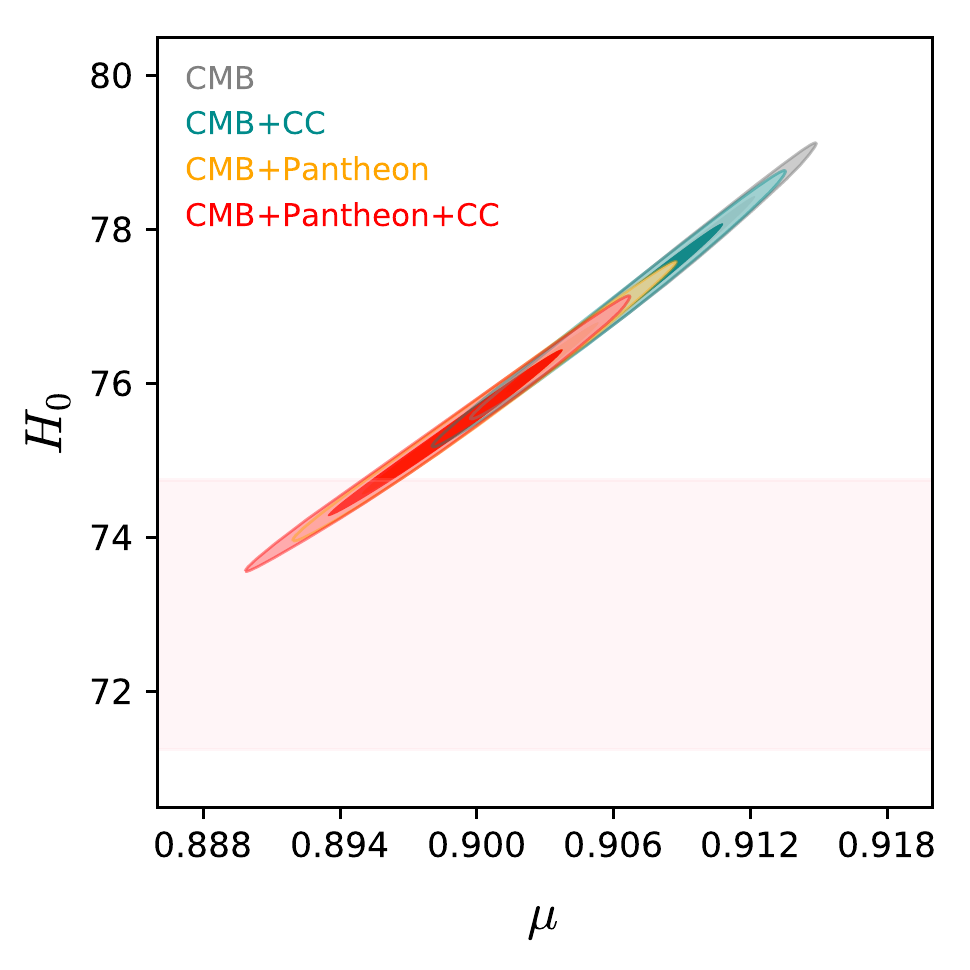}
\includegraphics[width=0.35\textwidth]{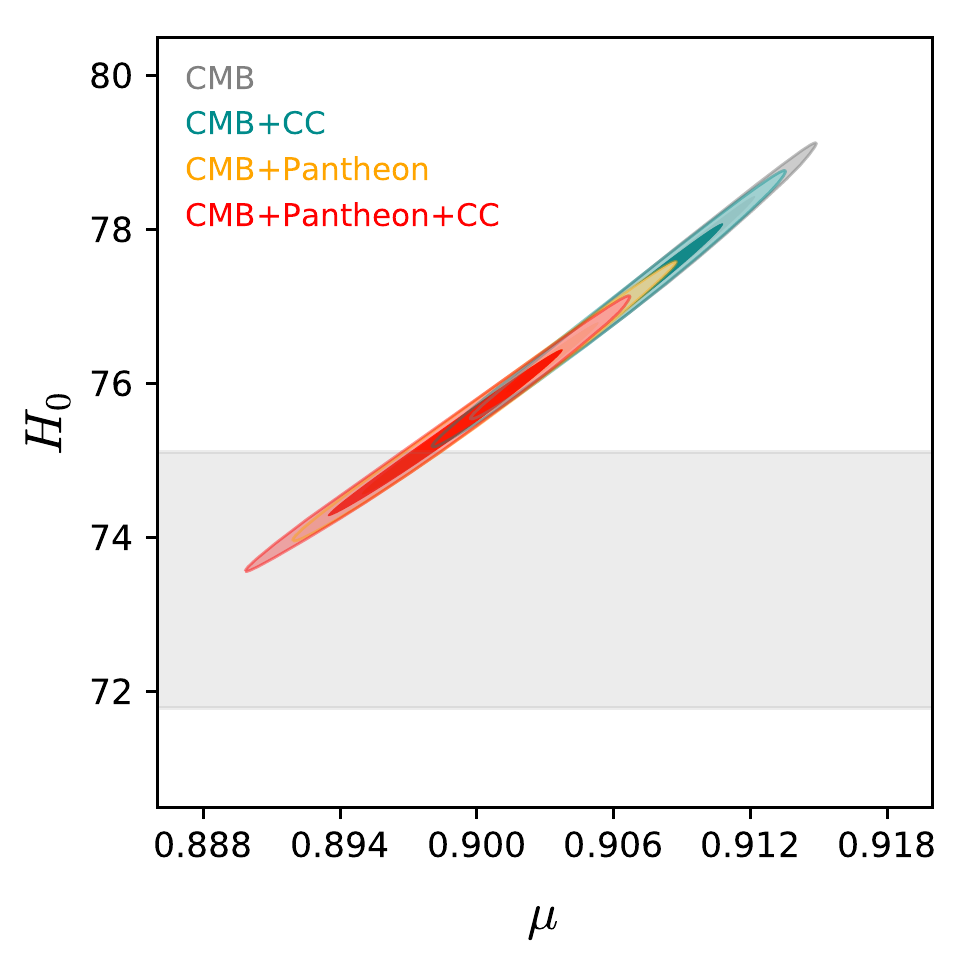}
\caption{The figure showing the plane ($\mu$, $H_0$) for the observational datasets CMB, CMB+CC, CMB+Pantheon, CMB+Pantheon+CC, indicates that the tension on $H_0$ can be  alleviated. In the left panel the pink shaded region stands for the $H_0$ band as reported by Riess et al. 2016 ($H_0 = 73.24 \pm 1.74$ km/sec/Mpc) \cite{Riess:2016jrr} 
while in the right panel, the grey shaded region stands for the $H_0$ band as reported by Riess et al. 2018 ($H_0 = 73.48 \pm 1.66$ km/s/Mpc) \cite{R18}.  }
\label{fig-tension}
\end{figure*}

\section{Observational data and the results}
\label{sec-data}

Here, we present the observational data used to fit the unified cosmological model and the methodology of the fitting technique.  

\begin{itemize}

\item CMB: The data from cosmic microwave background (CMB) are an effective observational probe for analyzing the cosmological models. We
consider the CMB temperature and polarization anisotropies along with their cross-correlations from Planck 2015 \cite{Adam:2015rua}. In particular, we include the 
combinations of high- and low-$\ell$ 
TT likelihoods in the multiple range $2\leq \ell \leq 2508$ as well as the combinations  of the high- and low-$\ell$ polarization likelihoods \cite{Aghanim:2015xee}.

\item Pantheon: We include the Pantheon sample \cite{Scolnic:2017caz} from the Supernovae Type Ia (SNIa). The pantheon sample is the most latest compilation 
of the the SNIa data comprising of 1048 data in the redshift range $z \in [0.01, 2.3]$ \cite{Scolnic:2017caz}. 
    
\item CC: Lastly, we consider  the Hubble parameter measurements from the cosmic chronometers (CC). The cosmic chronometers are actually some special kind of galaxies which are most massive and passively evolving. We refer to Ref. \cite{Moresco:2016mzx} for a detail reading on the methodology and the motivation to choose CC to measure the Hubble parameter values at different redshifts. Here, in this work, we 
consider 30 measurements of the Hubble parameter values distributed in the redshift range $0< z < 2$ \cite{Moresco:2016mzx}.

\item R18:  We consider a measurement of the  Hubble constant yielding $73.48 \pm 1.66$ km/s/Mpc by Riess et al. 2018 \cite{R18}. 
    
\end{itemize}

Now, to extract the constraints on the free and derived parameters of this unified cosmological  scenario, we use the Markov chain Monte Carlo package \texttt{cosmomc} \cite{Lewis:2002ah, Lewis:1999bs}, a fastest algorithm for the cosmological data analysis which also includes the support for the Planck 2015 likelihood code \cite{Aghanim:2015xee} (see the publicly available code here \url{http://cosmologist.info/cosmomc/}). We mention that \texttt{cosmomc} is already equipped with the Gelman-Rubin statistics \cite{Gelman-Rubin}. A part of our analysis includes the modifications of the CAMB. The modifications of the CAMB for this new unified dark fluid, we need to first modify the core codes, namely, equations.f90 and modules.f90, that means we need to modify both the background equations and the perturbation equations. These modifications enable us to obtain the background solution $\rho_u$, and perturbations solutions $\delta_u$, $\theta_u$. Further, using the codes, namely,  camb.f90, cmbmian.f90, and modules.f90, we calculate the output of the CMB temperature and matter power spectra. 
In connection with the fitting analysis, perhaps, we must mention that although the cosmological parameters for Planck 2018 are already published recently \cite{Aghanim:2018eyx}, however, the likelihood code is not public yet. Thus, we continue our analysis with the Planck 2015 data. Maybe the comparisons between the cosmological parameters obtained from Planck 2015 and Planck 2018 will be worth for a better understanding of the model.  So, in summary, we consider the following parameters space $\mathcal{P} \equiv \{\Omega_b h^2, 100 \theta_{MC}, \tau, n_s, {\rm{ln}}(10^{10} A_s), \mu\}$ and the priors on these parameters are enlisted in Table \ref{tab:priors}.  To begin with a robust observational analyses, initially we fit the model using the background data only, that means with CC, Pantheon and their combined data Pantheon+CC. In Table \ref{tab:background} we present their observational constraints at 68\% and 95\% CL. However, we note that the constraints achieved for CC, Pantheon and Pantheon+CC are not so stringent compared to the constraints when CMB data are added to the individual background data. This means, when the CMB data are added to these individual dataset, namely, CC, Pantheon and Pantheon+CC, the free and derived parameters of the unified model are significantly improved. This is clear if we compare the one-dimensional posterior distributions in Fig. \ref{fig:background}. Since from Fig. \ref{fig:background} one can clearly  visualize the significant constraining power of the CMB data compared to the background datasets, threfore, we shall be mainly interested on the constraints of the unified model in presence of the CMB data. In what follows we describe the observational datasets and their results. 
The first set we consider is the following (labeled as \texttt{Set I})
\begin{figure*}
\includegraphics[width=0.69\textwidth]{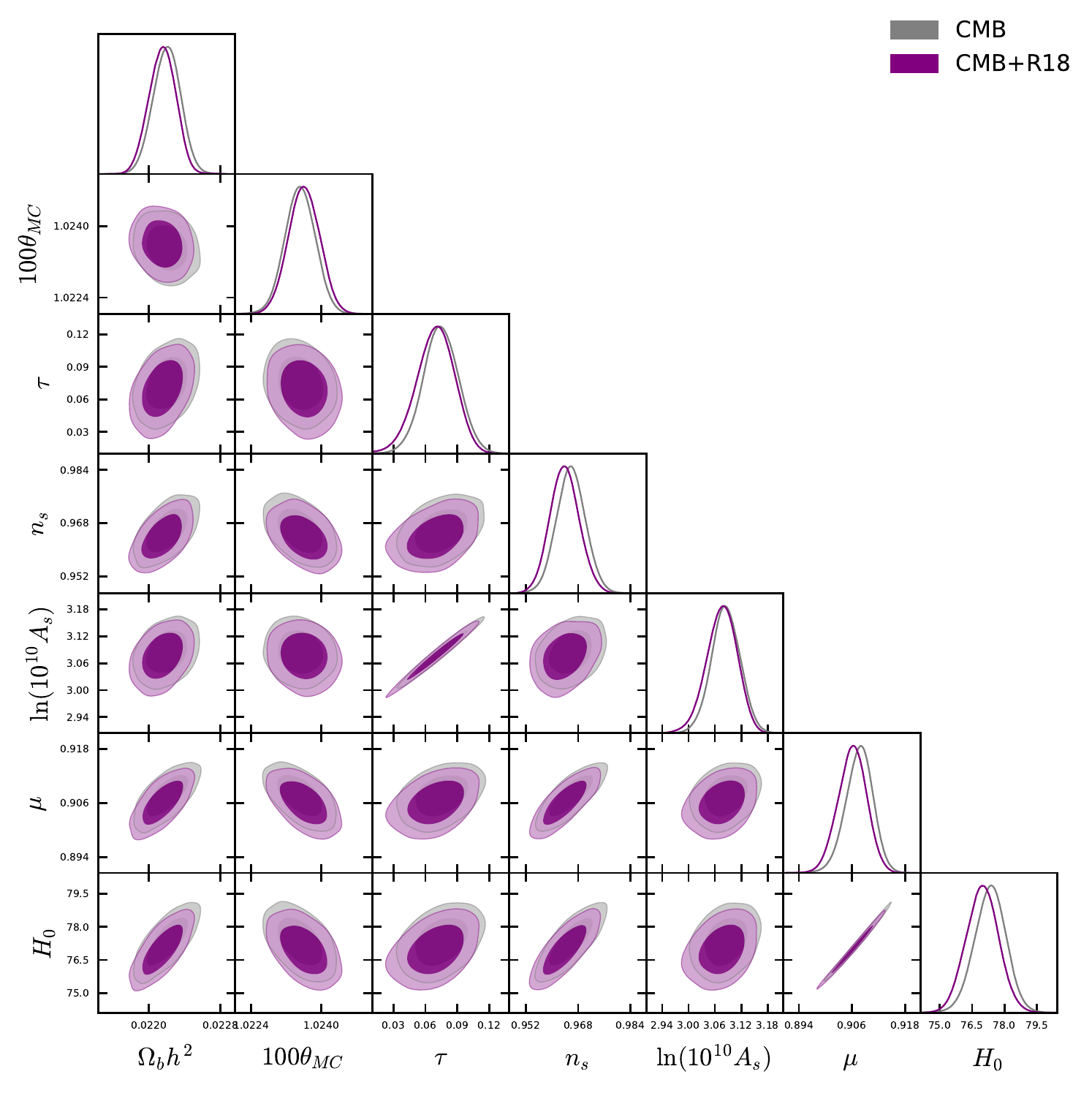}
\caption{The figure compares the observational constraints on the free and derived parameters of the unified dark model obtained from CMB and CMB+R18. One can clearly visualize that the addition of R18 data \cite{R18} does not found to improve the constraints obtained from CMB alone. }
\label{fig-compare1}
\end{figure*} 
\begin{figure*}
\includegraphics[width=0.69\textwidth]{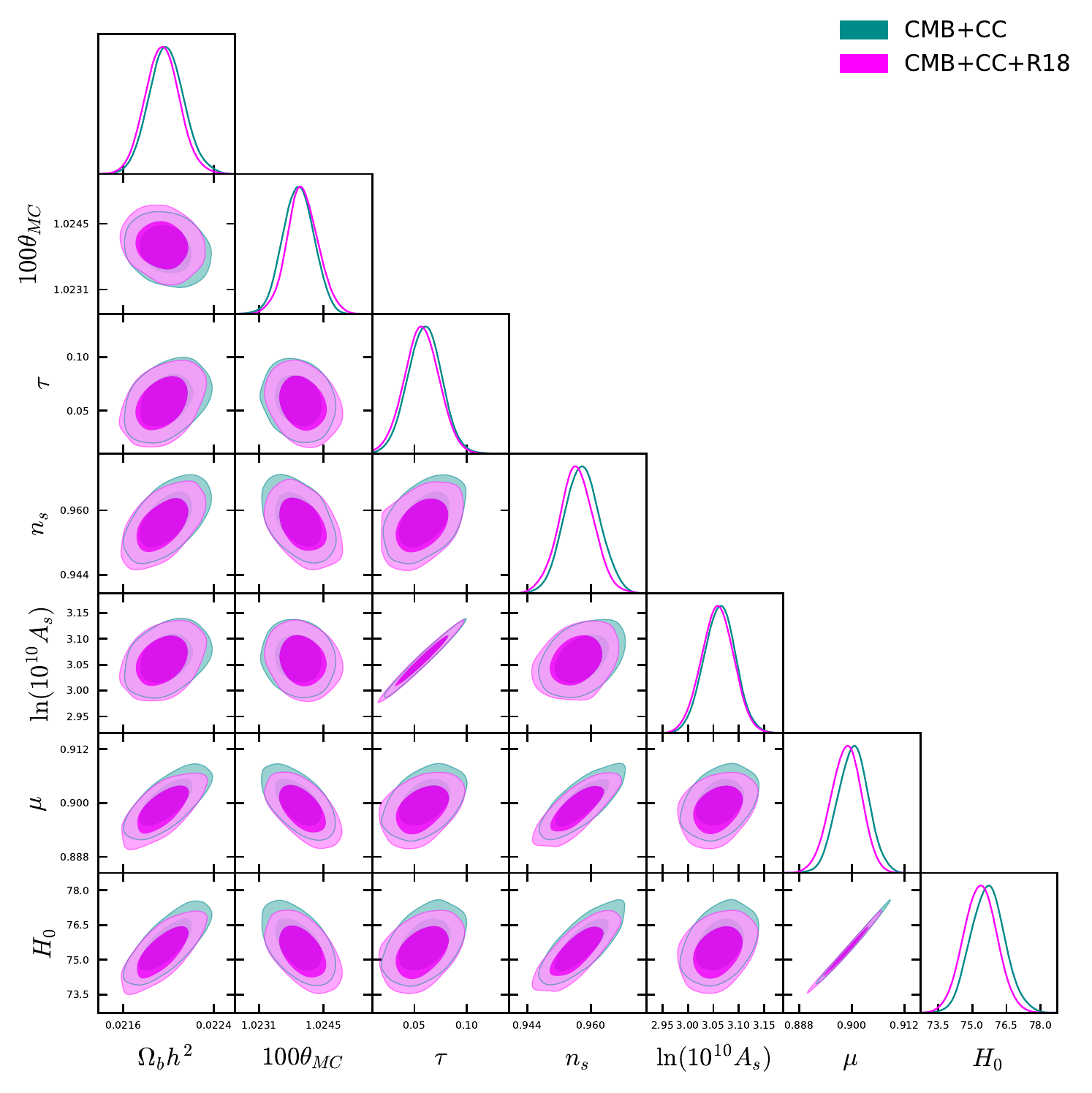}
\caption{The figure compares the observational constraints on the free and derived parameters of the unified dark model obtained from CMB+CC and CMB+CC+R18. Similar to Fig. \ref{fig-compare1}, here too we observe that the constraints from these datasets are almost same. }
\label{fig-compare2}
\end{figure*} 
\begin{figure*}
\includegraphics[width=0.69\textwidth]{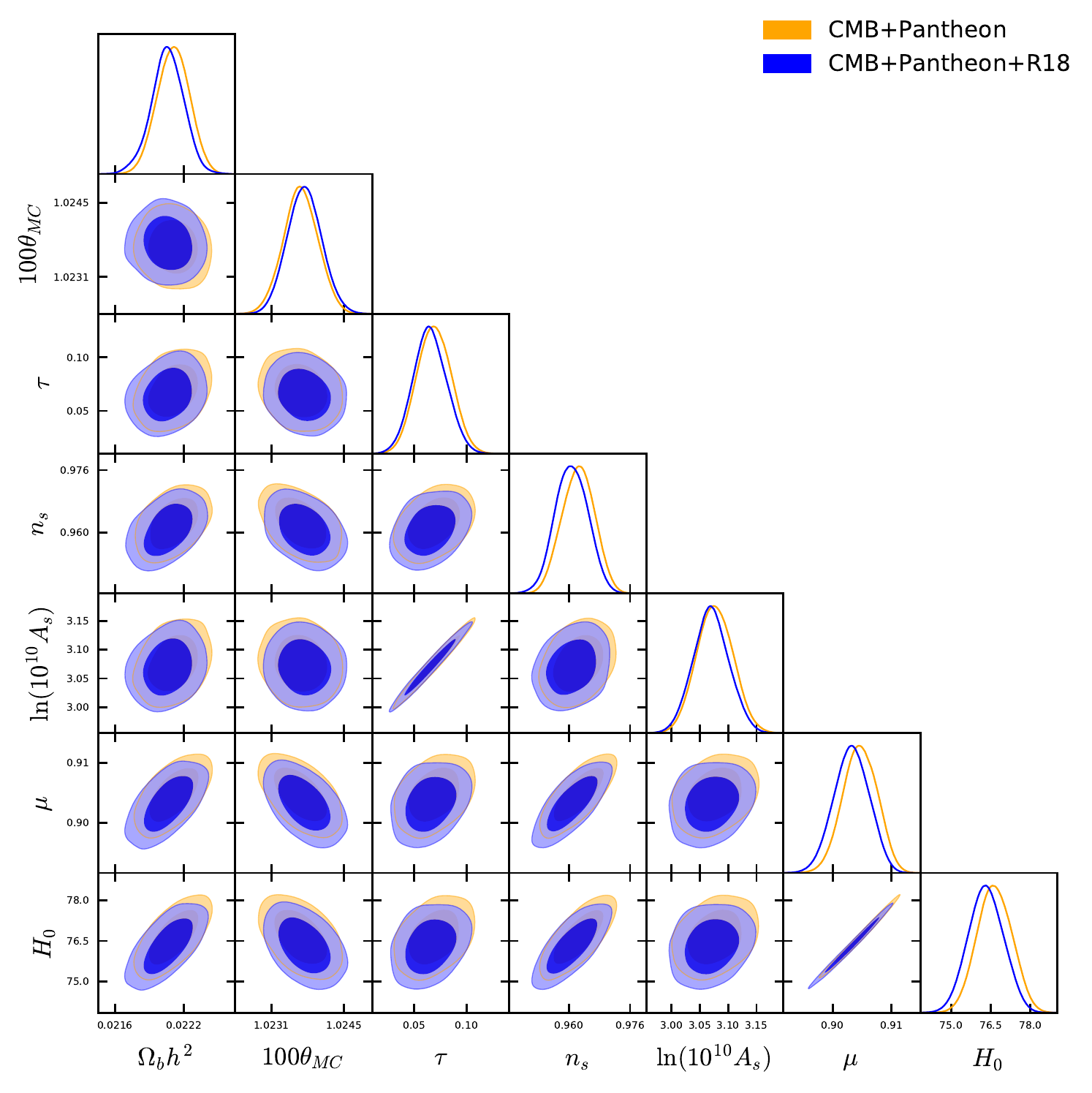}
\caption{The figure compares the observational constraints on the free and derived parameters of the unified dark model obtained from CMB+Pantheon and CMB+Pantheon+R18. As one can see there is almost no changes between the constraints from these two datasets. } 
\label{fig-compare3}
\end{figure*} 
\begin{figure*}
\includegraphics[width=0.69\textwidth]{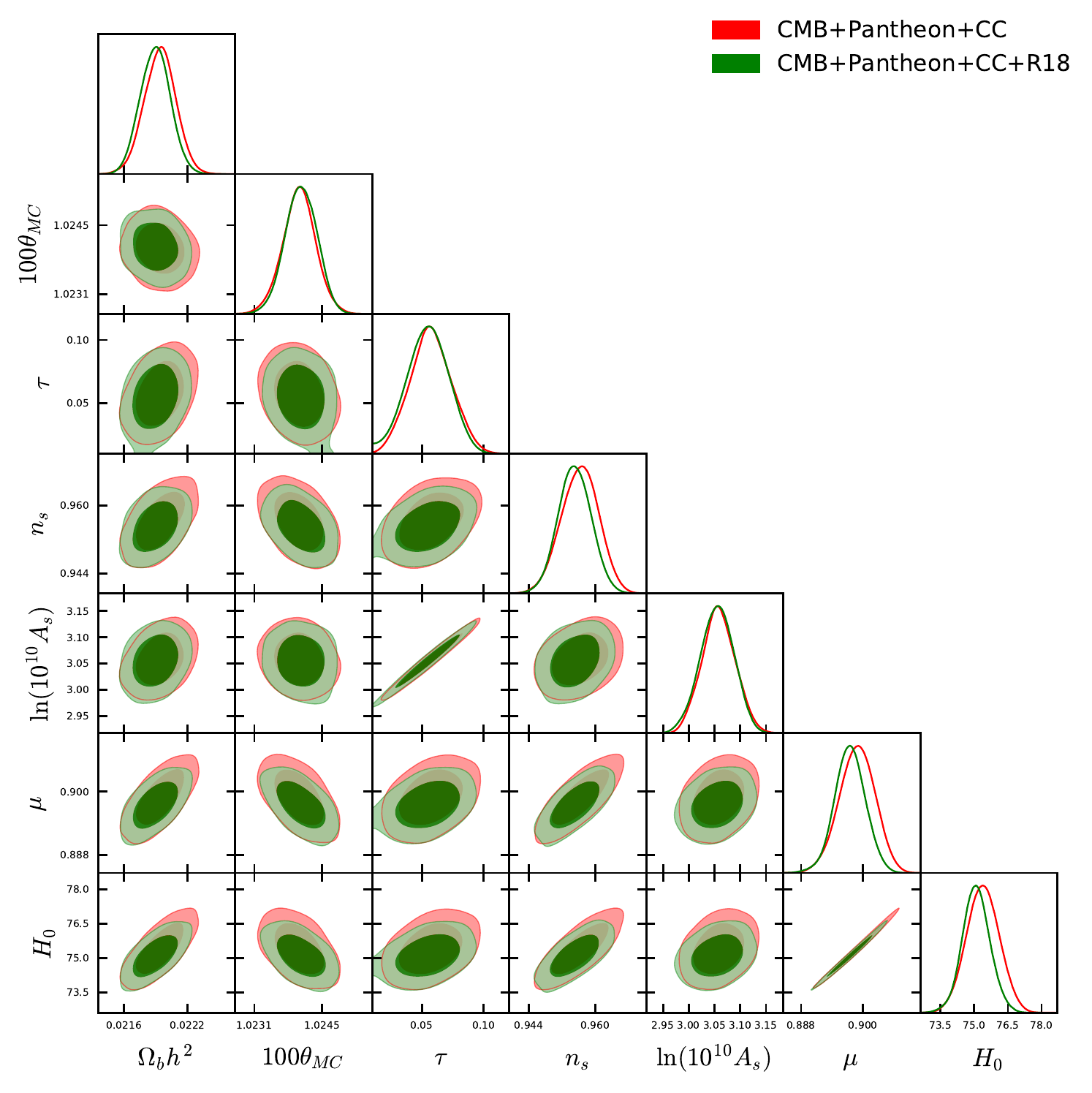}
\caption{The figure compares the observational constraints on the free and derived parameters of the unified dark model obtained from CMB+Pantheon and CMB+Pantheon+R18. Our observation remains same what we have seen in Fig. \ref{fig-compare1}, \ref{fig-compare2} and \ref{fig-compare3}. }
\label{fig-compare4}
\end{figure*} 
\begin{itemize}
\item CMB

\item CMB+CC

\item CMB+Pantheon

\item CMB+Pantheon+CC
\end{itemize}
and the second set of datasets is the following (labeled as \texttt{Set II})
\begin{itemize}
\item CMB+R18 
\item CMB+CC+R18 
\item CMB+Pantheon+R18 
\item CMB+Pantheon+CC+R18
\end{itemize}
Now, using these two different set of datasets, we have constrained the model parameters. Let us present the observational constraints for both  \texttt{Set I}
and \texttt{Set II} in a  systematic way. \\

\texttt{Set I}: The observational constraints for the first set of datasets are given in  
Table \ref{tab:results1} where we have shown the constraints at 68\% and 95\% CL. Further, in order to understand the graphical behaviour of the parameters,     
in Fig. \ref{fig-2D} we show the 1-dimensional marginalized posterior distributions for the parameters of this cosmological scenario as well as we show the 2D contour plots between several combinations of the model parameters at 68\% and 95\% CL.  
From Table \ref{tab:results1}, one can clearly see that  CMB data alone return a very higher value of the Hubble constant, $H_0 = 77.33_{-    0.73}^{+    0.71}$ at 68\% CL ($H_0 =  77.33_{- 1.48}^{+ 1.41}$ at 95\% CL). This estimation is very large than the Planck's ($\Lambda$CDM) based estimation \cite{Ade:2015xua}. The addition of CC slghtly lowers the estimation of $H_0$, however, overall the constraints on $H_0$ are compatible to the local estimation by Riess et al. 2016 ($H_0 = 73.24 \pm 1.74$ km/sec/Mpc) \cite{Riess:2016jrr} and by Riess et el. 2018 ($H_0 = 73.48 \pm 1.66$ km/s/Mpc)  \cite{R18}. In fact, looking at the posteriors of the model parameters as well as the two dimensional contour plots depicted in Fig. \ref{fig-2D},   
one can see that CMB data alone and CMB+CC dataset return almost similar fit to the model parameters. 
Now the adition of Pantheon to CMB has some visible effects on the model parameters. 
In this case we see that although $H_0$ takes higher values ($H_0 = 75.68_{-0.76}^{+    0.74}$, at 68\% CL, CMB+Pantheon) compared to the $\Lambda$CDM based Planck's report \cite{Ade:2015xua}, but compared to the constraints from CMB alone datset summarized in the second column of Table \ref{tab:results1}, $H_0$ takes lower values. And moreover the deviation in the mean value of the Hubble constant $H_0$ defined by $\Delta H_0 = |H_0 ({\rm CMB}) - H_0 ({\rm CMB+Pantheon})| = 1.65$ km/s/Mpc, is not insignificant indeed. Due to the higher values of $H_0$, the tension on $H_0$ is released for this dataset. Concerning the  final dataset CMB+Pantheon+CC, we do not find anymore constraining power of CC that may exceed the constraining power of CMB+Pantheon. 
This becomes clear if one looks at the   Fig. \ref{fig-2D}, speicifically the 1-dimensional posterior distributions and the 2-dimensional contour plots between CMB+Pantheon and CMB+Pantheon+CC. Thus, overall, 
since for all the observational datasets,  $H_0$ assumes larger values, so one can notice that for this unified cosmological model, the tension on $H_0$ coming from the global \cite{Ade:2015xua} and local measurements \cite{Riess:2016jrr, R18} is released. In Fig. \ref{fig-tension} we have clearly displayed this issue considering the values from Riess et al 2016 (left panel of Fig. \ref{fig-tension})  and Riess et al. 2018 (right panel of Fig. \ref{fig-tension}). This might be considered to be an interesting property of the present unified fluid.  A very recent work by Riess et al. \cite{Riess:2019cxk} pointed out that  local estimation of $H_0$ actually increases 
($H_0 = 74.02 \pm 1.42$ km/s/Mpc) compared to two earlier reports  \cite{Riess:2016jrr, R18}. That means the present unified model  might be considered to be an excellent example to alleviate the $H_0$ tension along the similar lines of some earlier investigations \cite{DiValentino:2016hlg,Kumar:2017dnp,DiValentino:2017iww,Vagnozzi:2018jhn,Yang:2018euj,Yang:2018uae,Yang:2018qmz,Pan:2019jqh} that also solve the $H_0$ tension. 
Finally, we comment on the main free parameter $\mu$ of the present unified model. From Table \ref{tab:results1} we see that $\mu$ is around $0.9$ (since its erorr bars are so small) and recalling the graphical behaviour of the deceleration parameter for this unified model (see the left plot of Fig. \ref{fig-deceleration}), one can conclude that since $\mu \lesssim 0.9$, thus the transition from the past decelerating phase to the current accelerating phase happens at around $z \lesssim 0.6$ which is consistent according to the recent observational facts. However, at the end we mention that according to the observational fittings, $\mu$ has a strong positive correlation with $H_0$. \\

\texttt{Set II:} We now focus on the observational constraints on the model parameters after the inclusion of the local measurement of $H_0$ by Riess et al. 2018 \cite{R18} with the previous datasets (CMB, Pantheon, CC) in order to see how the parameters could be improved with the inclusion of this data point.  Since for this present unified model the estimation of $H_0$ from CMB alone is compatible with the local estimation of $H_0$   by Riess et al. 2018 \cite{R18}, thus, we can safely add both the datasets to see whether we could have something interesting.  Following this, we perform another couple of tests after the inclusion of R18. 
The observational results on the model parameters are summarized in Table \ref{tab:results2}. However, comparing the observational constraints reported in Table \ref{tab:results1} (without R18 data) and Table \ref{tab:results2} (with R18), one can see that the inclusion of R18 data \cite{R18} does not seem to improve the constraints on the model parameters. In fact, the estimation of the Hubble constant $H_0$ remains almost similar to what we found in Table \ref{tab:results1}.  In order to be more elaborate in this issue, we have compared the observational constraints on the model parameters before and after the inclusion of R18 to other datasets.  In Figs. \ref{fig-compare1} (CMB versus CMB+R18), \ref{fig-compare2} (CMB+CC versus CMB+CC+R18), \ref{fig-compare3} (CMB+Pantheon versus CMB+Pantheon+R18) and  \ref{fig-compare4} (CMB+Pantheon+CC versus CMB+Pantheon+CC+R18) we have shown the comparisons which prove our claim. One can further point out that the strong correlation between the parameters $\mu$ and $H_0$ as observed in Fig. \ref{fig-2D} still remains after the inclusion of R18 (see specifically the ($\mu$, $H_0$) planes in Figs. \ref{fig-compare1}, \ref{fig-compare2}, \ref{fig-compare3} and  \ref{fig-compare4}). The physical nature of $\mu $ does not alter at all. That means the correlation between $H_0$ and $\mu$ is still existing after the inclusion of R18 to the previous datasets, such as CMB, Pantheon, CC.  In addition to that since $\mu \lesssim 0.9$ according to all the observational datasets, thus, the transition from past decelerating era to current accelerating era occurs to be around $z \lesssim 0.6$, similar to what we have found with previous datasets (Table \ref{tab:results1}).

\begin{figure*}
\includegraphics[width=0.24\textwidth]{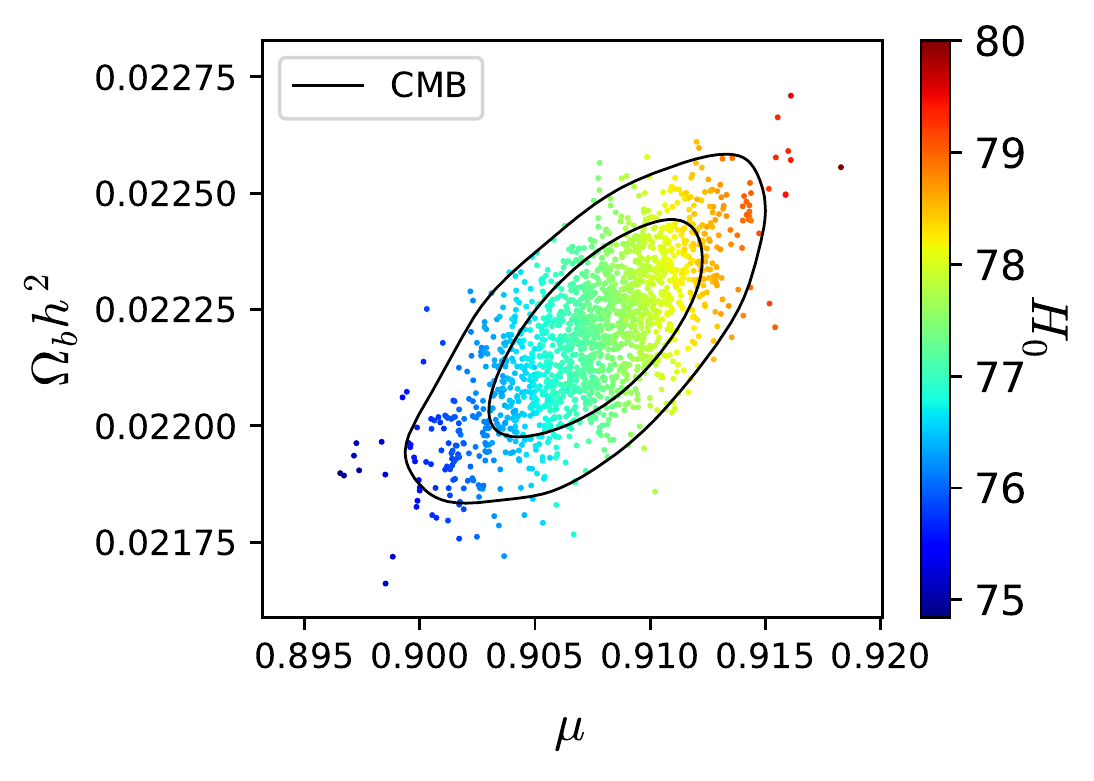}
\includegraphics[width=0.24\textwidth]{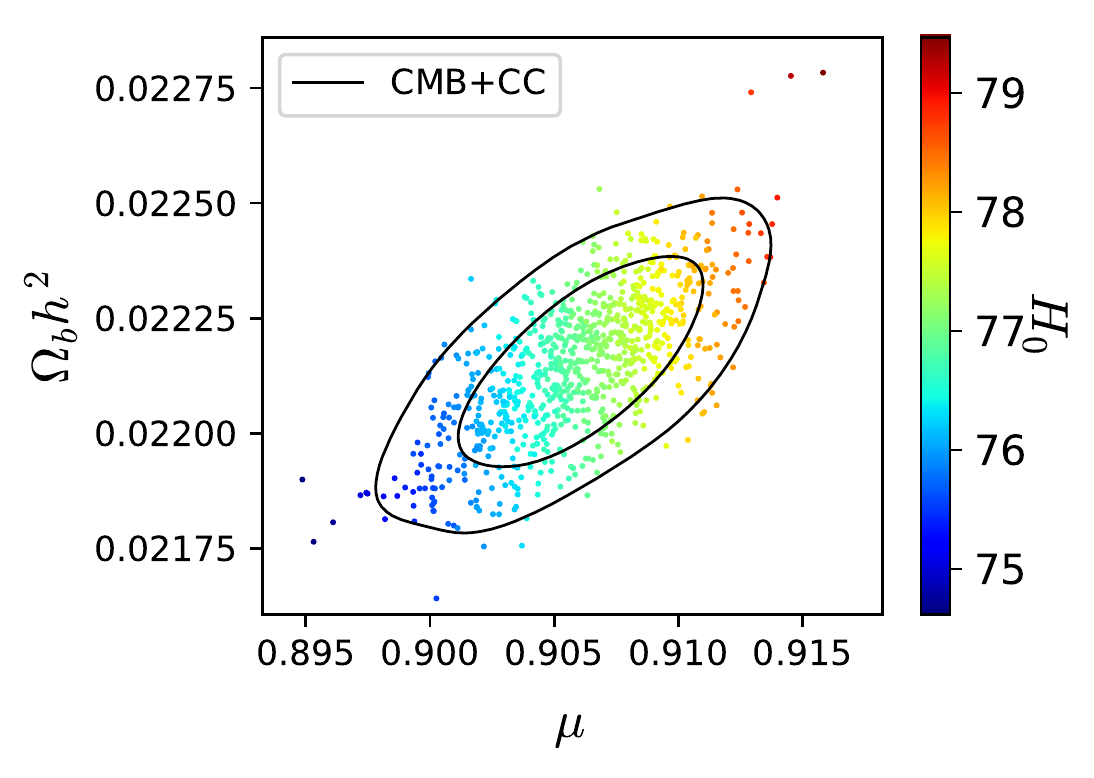}
\includegraphics[width=0.24\textwidth]{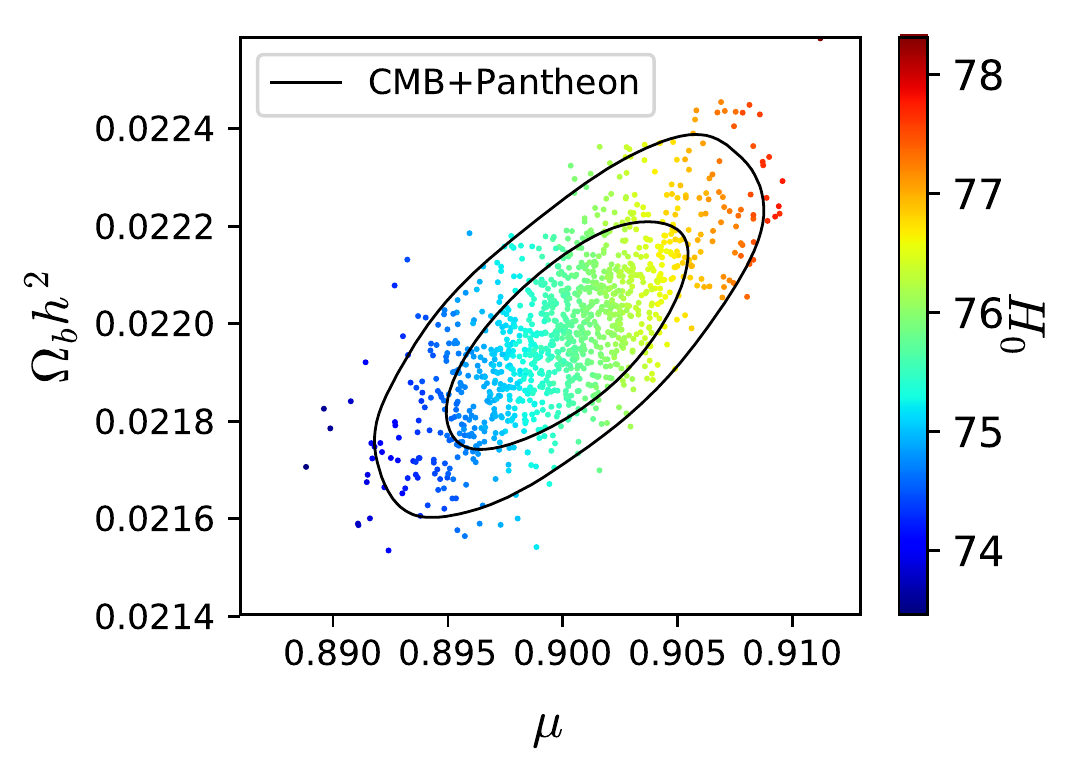}
\includegraphics[width=0.24\textwidth]{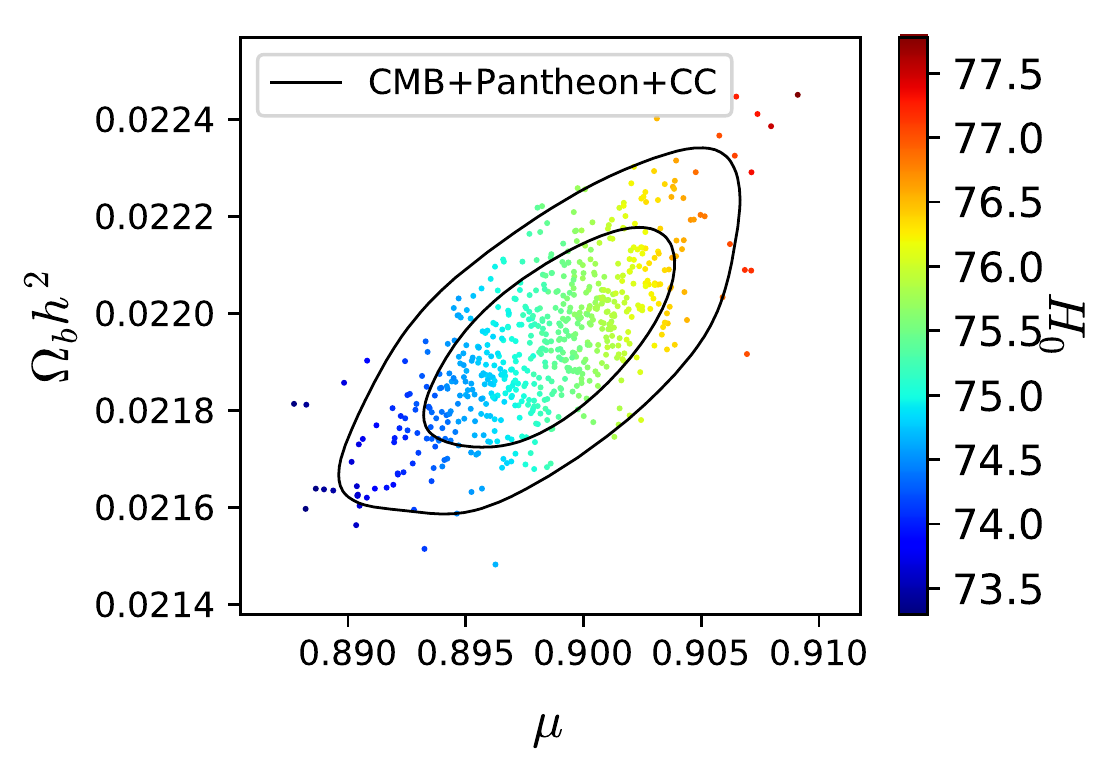}
\includegraphics[width=0.24\textwidth]{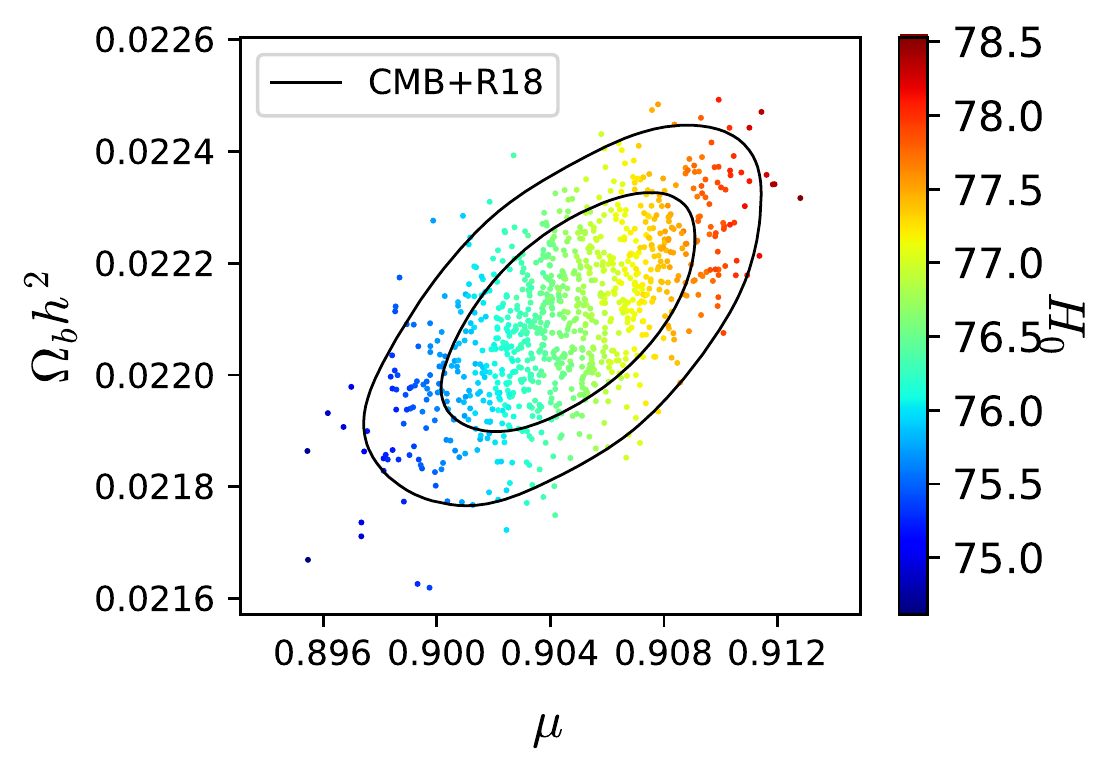}
\includegraphics[width=0.24\textwidth]{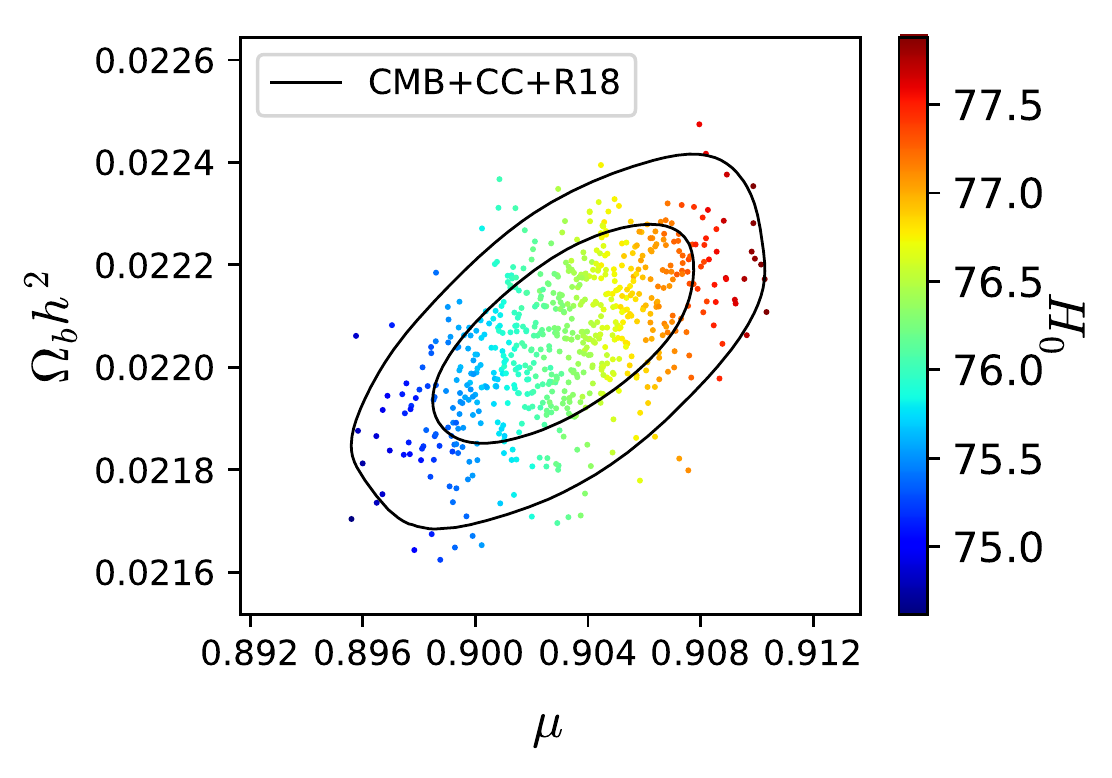}
\includegraphics[width=0.24\textwidth]{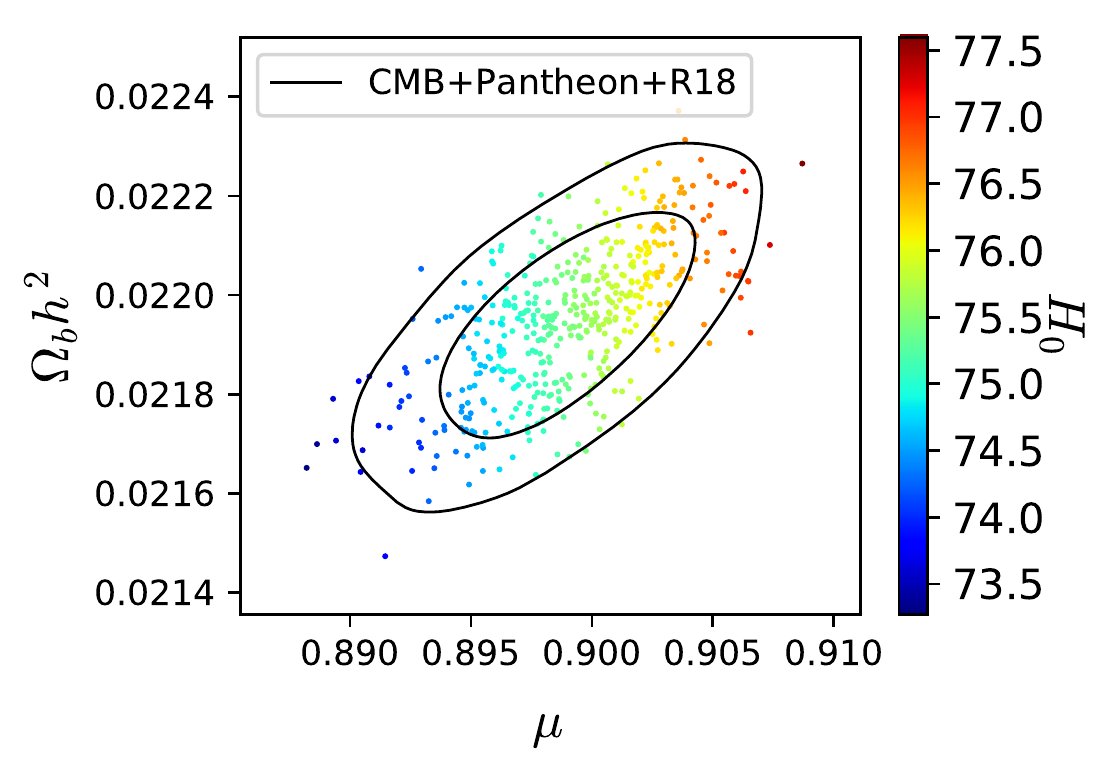}
\includegraphics[width=0.24\textwidth]{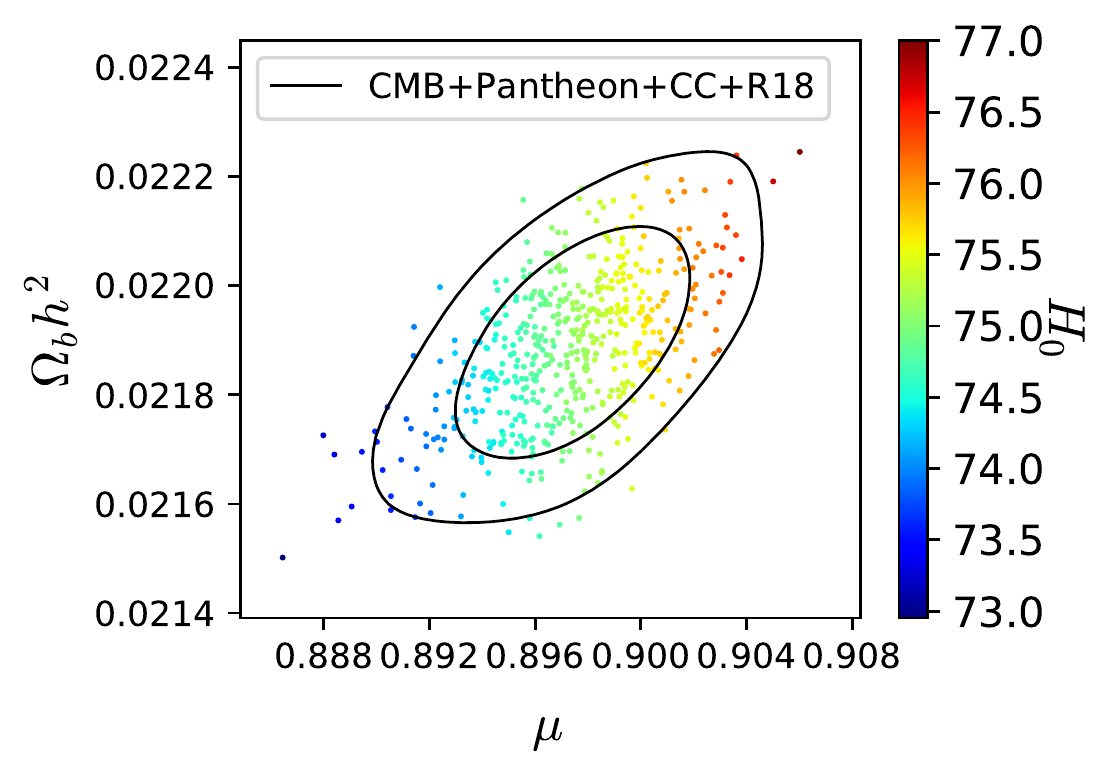}
\caption{We show the trend of parameters ($\mu$, $\Omega_b h^2$) for the present unified cosmological model for different values of the  $H_0$ parameter taken from the markov chain monte carlo chain for various observational data and their combinations. }
\label{fig:3DA}
\end{figure*}

\begin{figure*}
\includegraphics[width=0.24\textwidth]{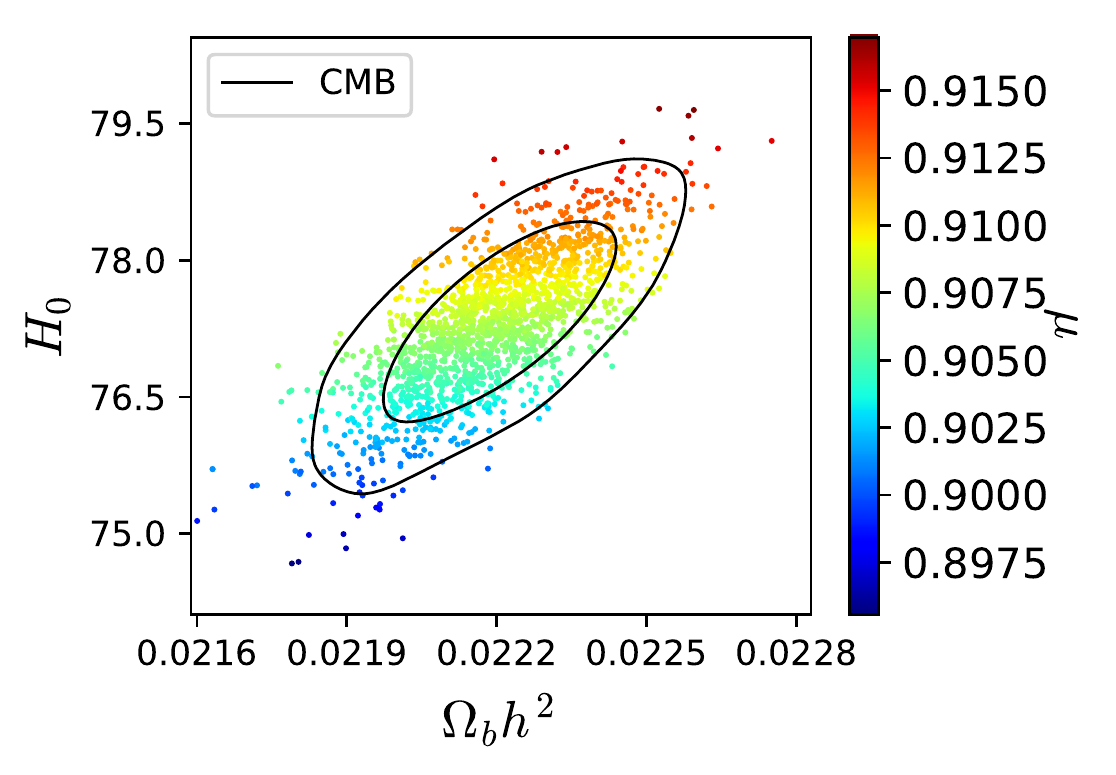}
\includegraphics[width=0.24\textwidth]{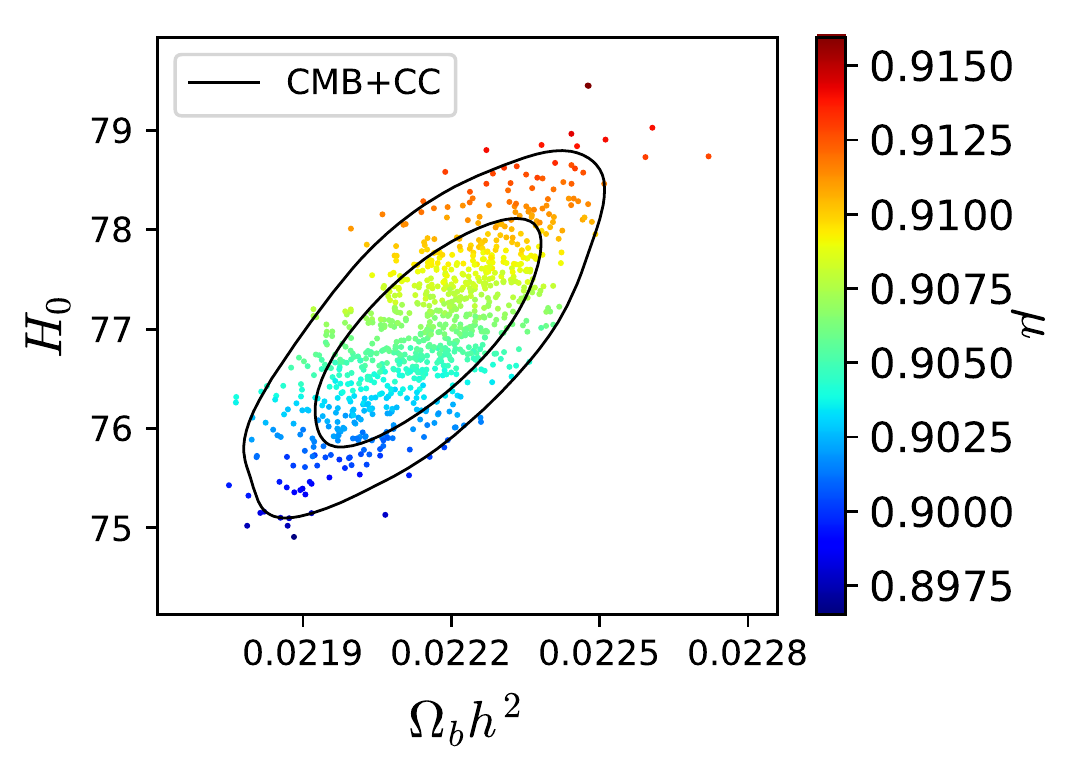}
\includegraphics[width=0.24\textwidth]{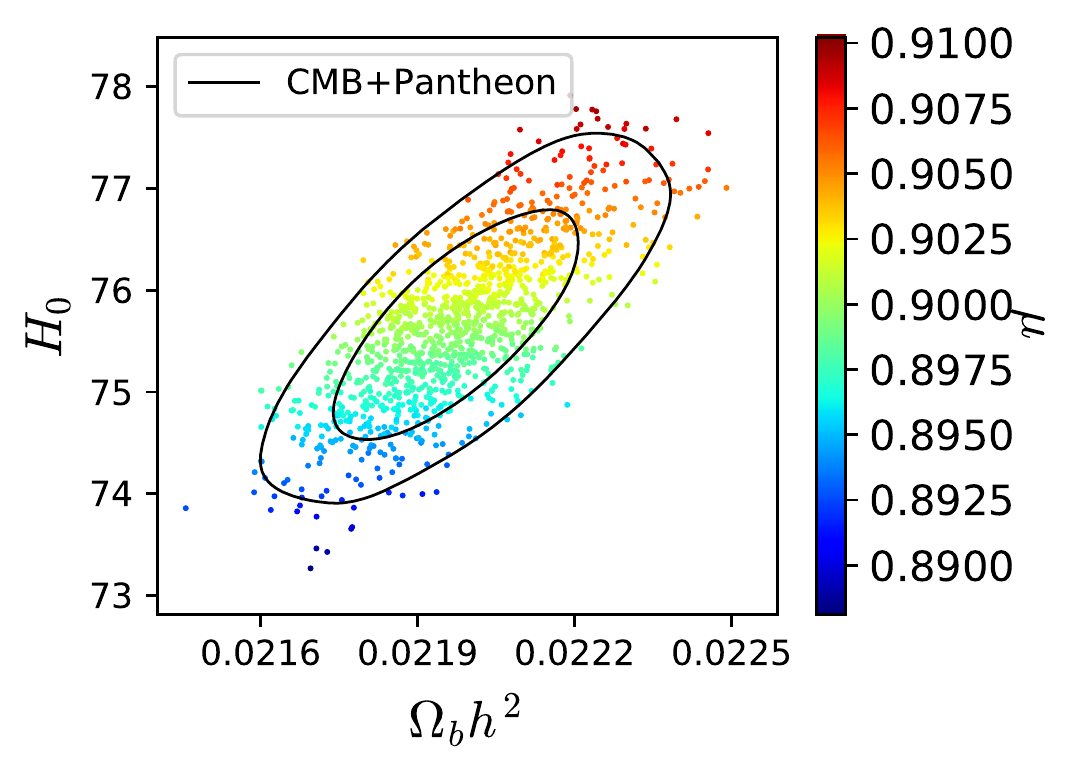}
\includegraphics[width=0.24\textwidth]{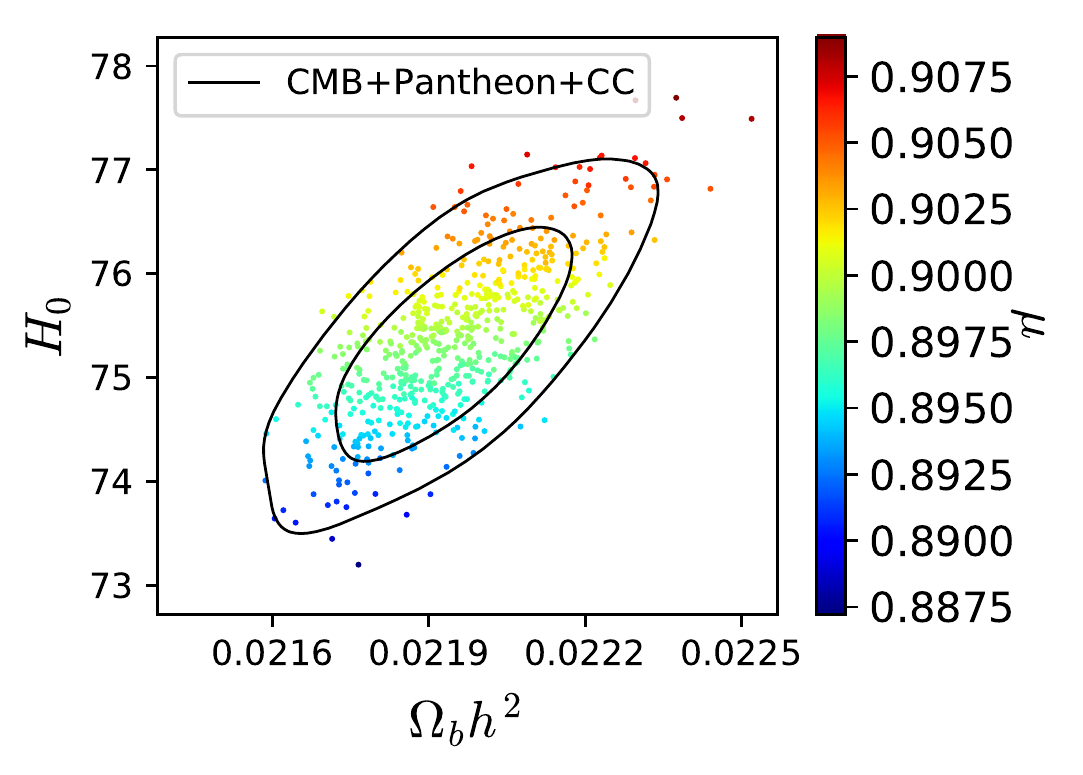}
\includegraphics[width=0.24\textwidth]{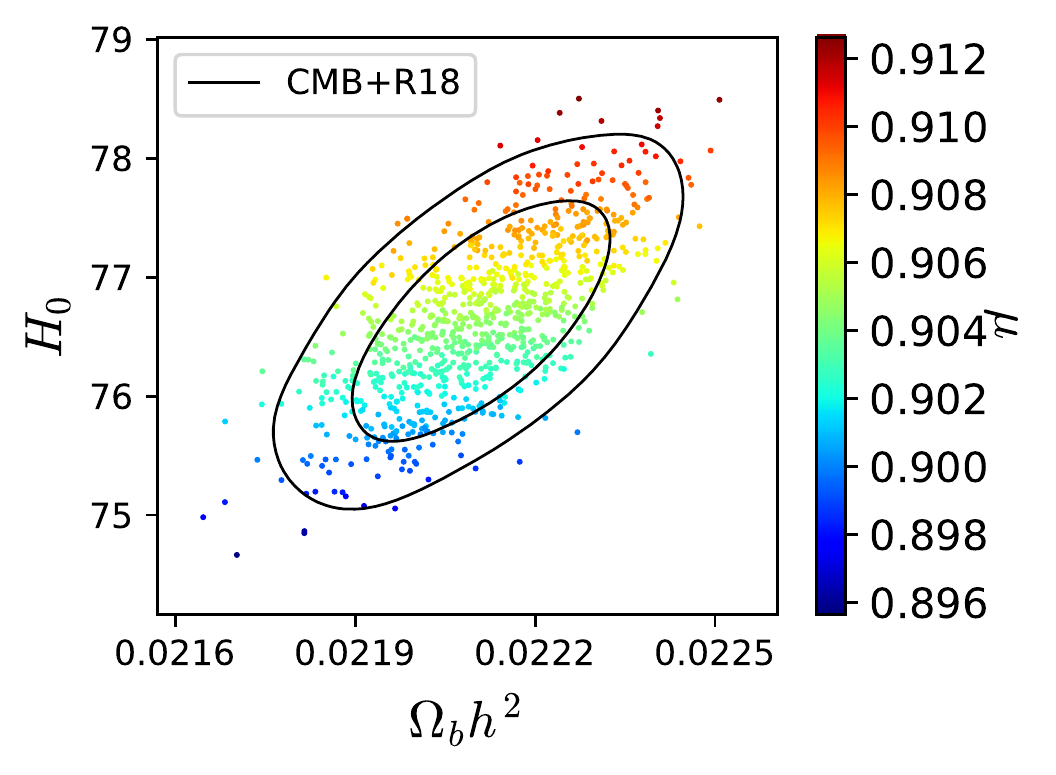}
\includegraphics[width=0.24\textwidth]{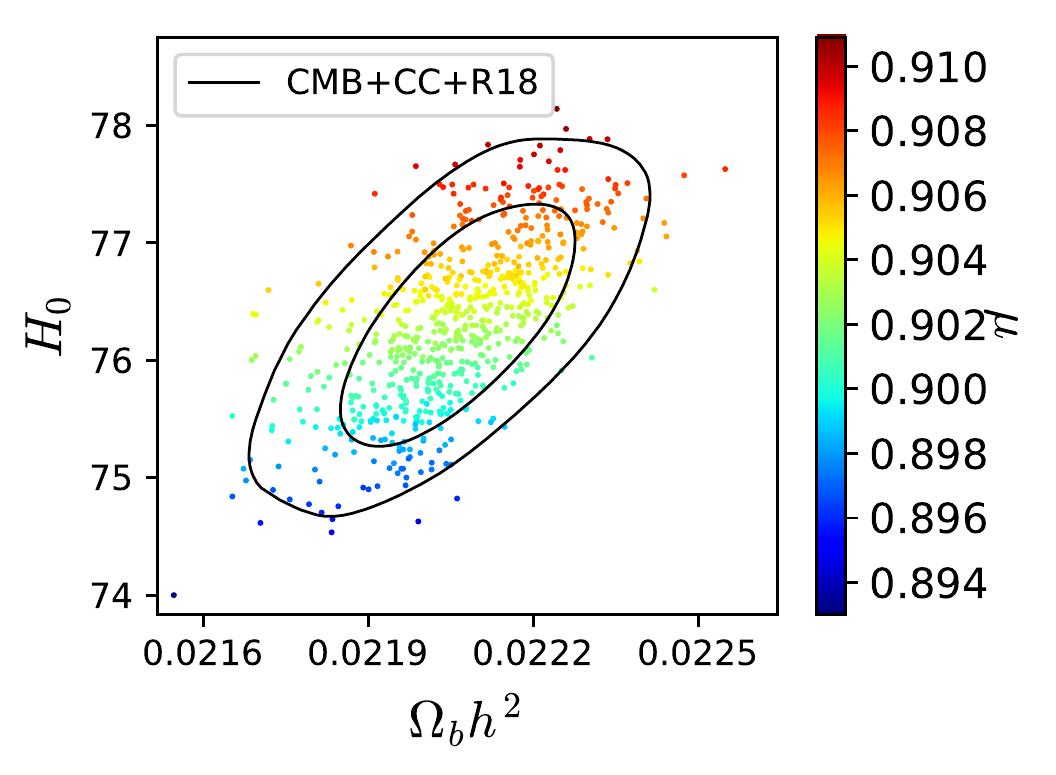}
\includegraphics[width=0.24\textwidth]{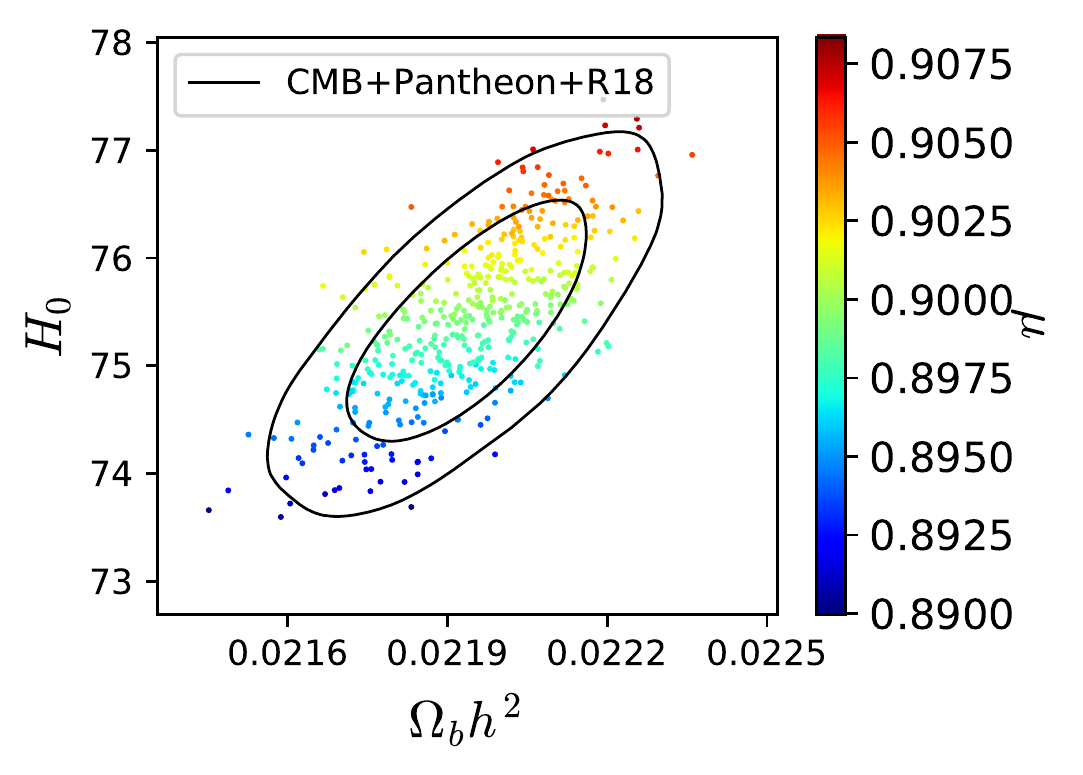}
\includegraphics[width=0.24\textwidth]{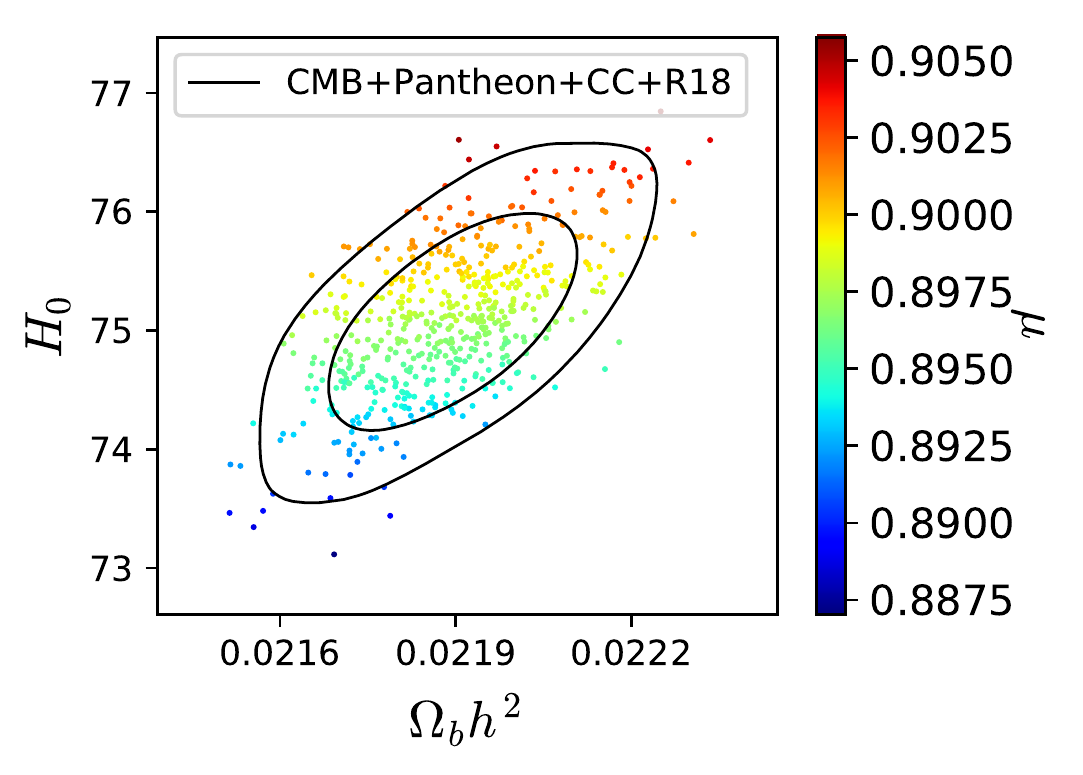}
\caption{We show the trend of parameters ($\Omega_b h^2$, $H_0$) for the present unified cosmological model  for different values of the  $\mu$ parameter taken from the markov chain monte carlo chain for various observational data and their combinations.}
\label{fig:3DB}
\end{figure*}

Before moving to the analysis of the model in the large scale of the universe we have shown two additional plots in Fig. \ref{fig:3DA} and Fig. \ref{fig:3DB} in which we have shown the trend of the model parameters. In particular, in Fig. \ref{fig:3DA} we have shown the  trend of the parameters ($\mu$, $\Omega_b h^2$) using the values of $H_0$ taken from the markov chain monte carlo (mcmc) sample for different observational data used in this work. In a similar fashion, in Fig. \ref{fig:3DB} we have described the trend of the other two free parameters ($\Omega_b h^2$, $H_0$) using different values of the parameter $\mu$ taken from the mcmc sample for different observational data. The figures \ref{fig:3DA}, and \ref{fig:3DB} provide with a clear image of the behaviour of the model parameters and their mutual dependence.

\begin{figure*}
\includegraphics[width=0.35\textwidth]{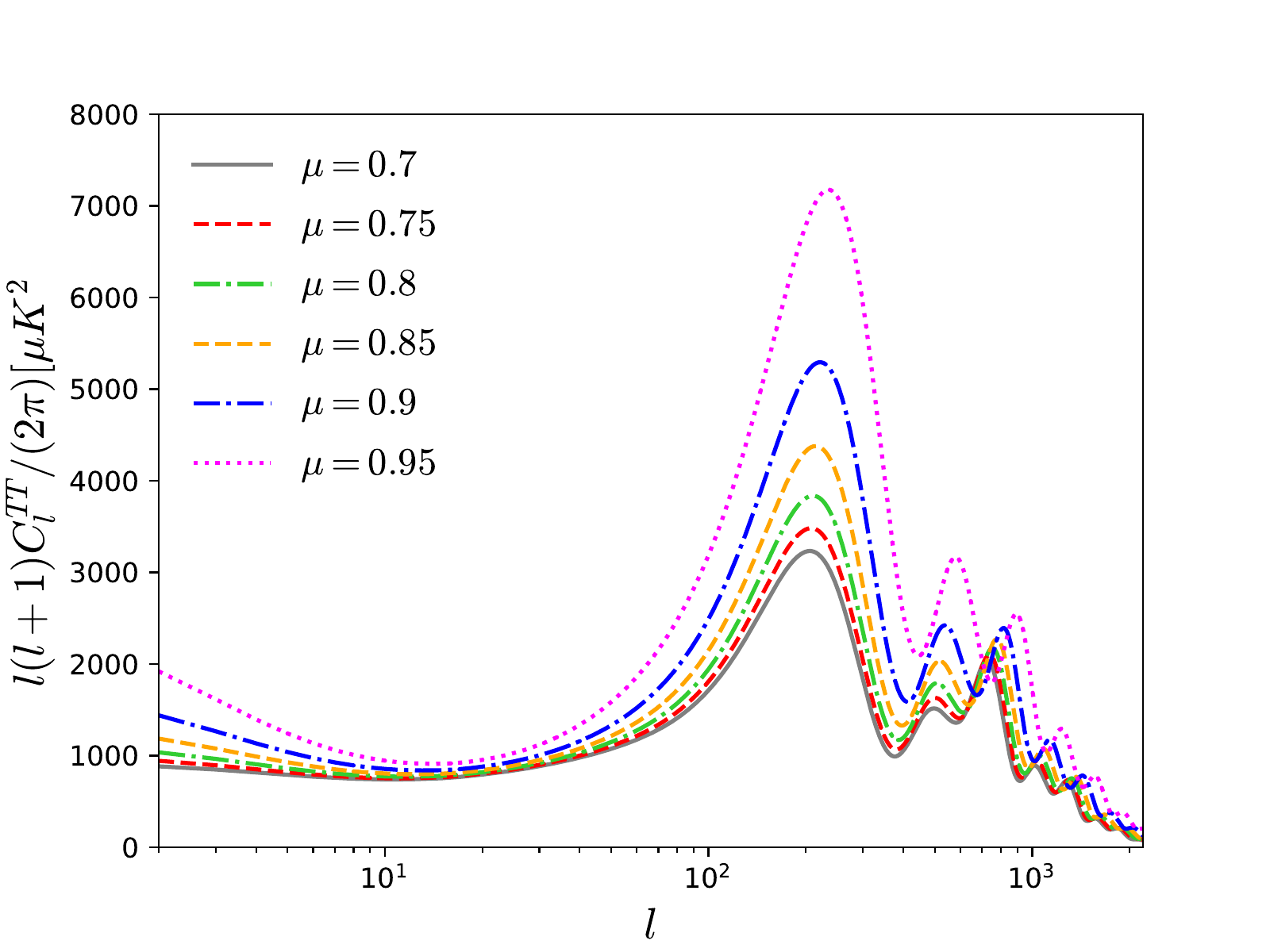}
\includegraphics[width=0.35\textwidth]{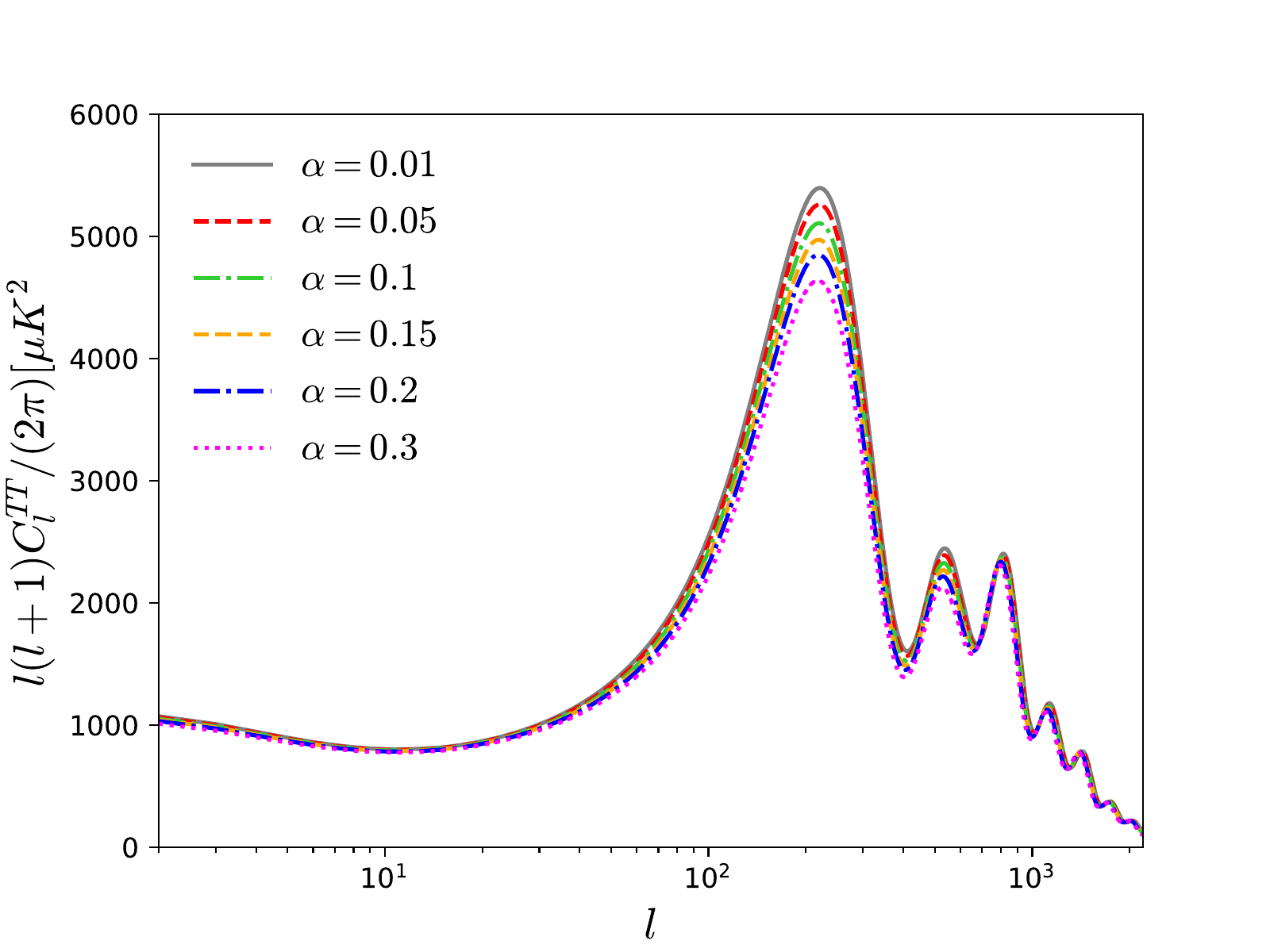}
\caption{We show and also present a qualitative comparison between the present unified cosmological model (left graph) and the generalized Chaplygin model (right graph) through the temperature anisotropy in the CMB TT spectra using different values of $\mu $ and $\alpha$. From the left and right graphs of this figure, one can clearly notice that the key parameter $\mu$ of the present unified model seems to be sensitive compared to the key parameter $\alpha$ of the GCG model. }
\label{fig-cmbpower}
\end{figure*}

\begin{figure*}
\includegraphics[width=0.35\textwidth]{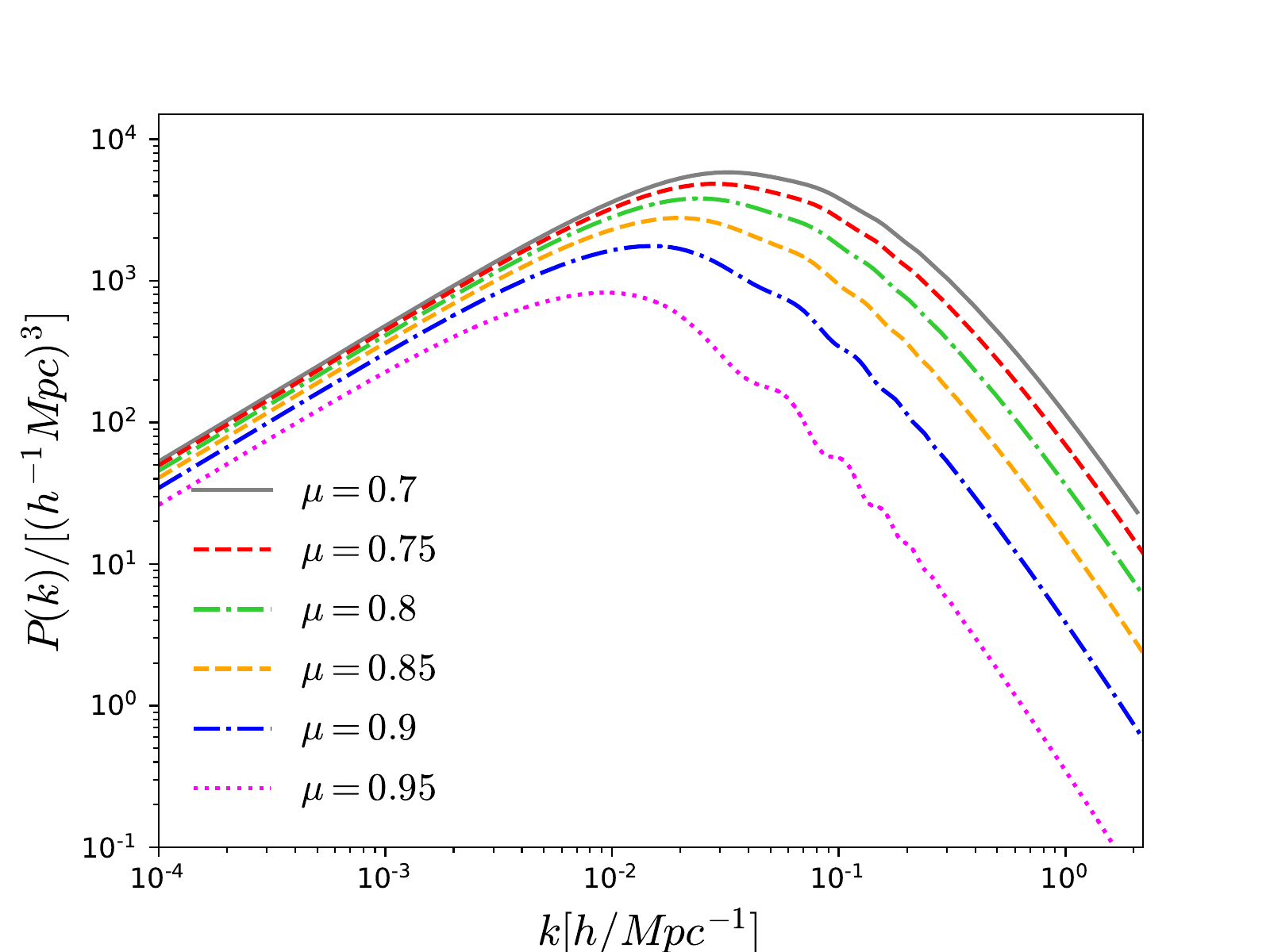}
\includegraphics[width=0.35\textwidth]{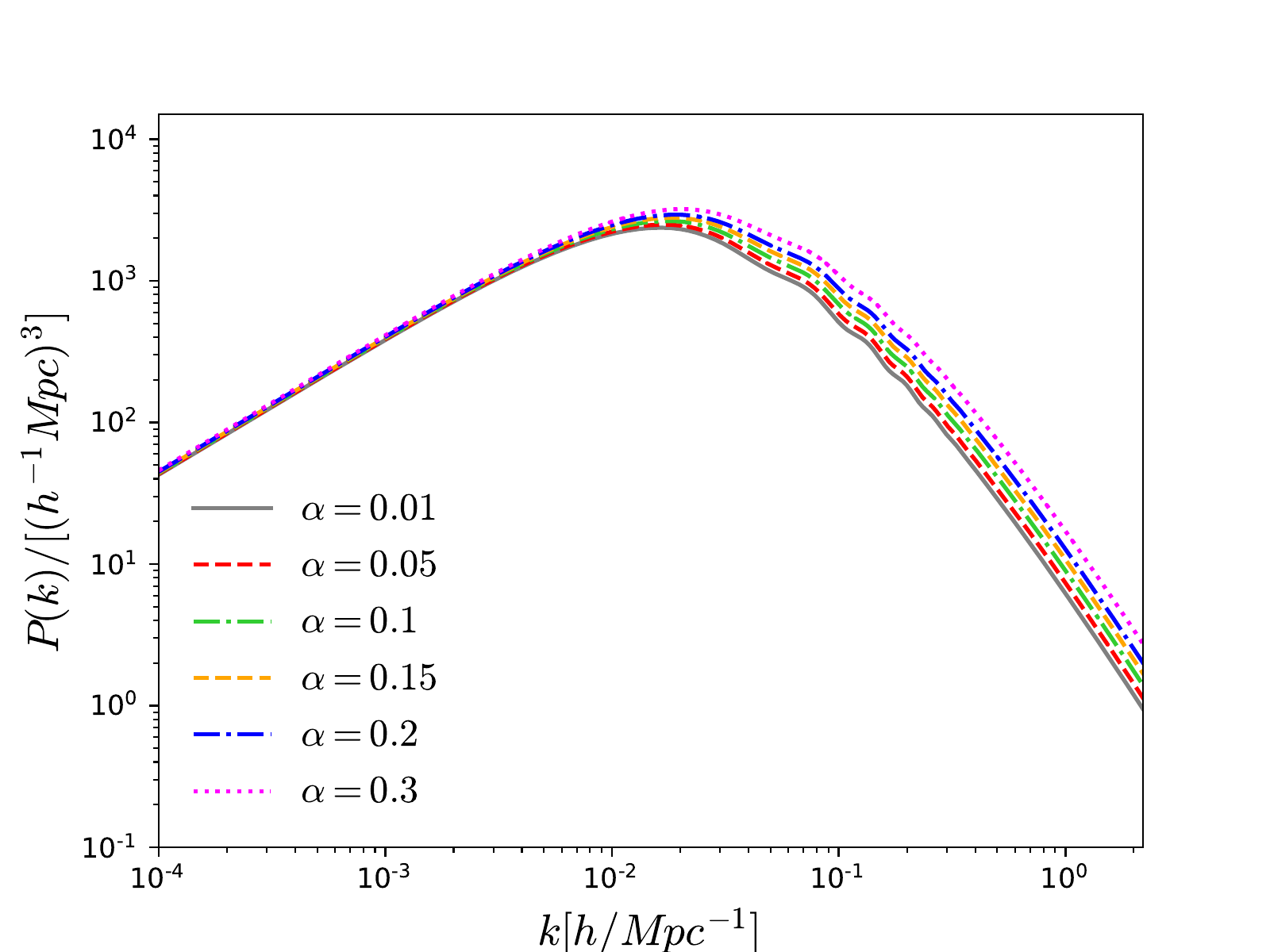}
\caption{We show and also present a qualitative comparison between the present unified cosmological model (left graph) and the generalized Chaplygin model (right graph) via matter power spectra using different values of $\mu $ and $\alpha$. Similar to the previous Fig. \ref{fig-cmbpower}, here too, we can clearly understand the sensitivity of the $\mu$ parameter compared to the $\alpha$ parameter. }
\label{fig-mpower}
\end{figure*}

We  continue our analysis by investigate the behaviour of the model in the large scale of the universe via the temperature anisotropy in the CMB TT spectra and the matter power spectra. Moreover, we have also made a comparison of the model with the generalized Chaplygin gas (GCG) model in order to understand how the models differ from one another. Thus, in order to do so,   
in the left graph of Fig. \ref{fig-cmbpower} we show the temperature anisotropy in the CMB TT spectra (left panel of Fig. \ref{fig-cmbpower}) for various values of the key parameter $\mu$ and at the same time in the right panel of Fig. \ref{fig-cmbpower}, we show the temperature anisotropy in the CMB TT spectra for the GCG model where we have used different values of $\alpha$. From this figure (Fig. \ref{fig-cmbpower}), one can notice that the parameter $\mu$ seems to be significant compared to the single parameter $\alpha$ of the GCG model. From the left graph of Fig. \ref{fig-cmbpower}, we see that with the increase of $\mu$, the height of the first acoustic peak in the CMB TT spectra increases which in contrary to the right graph of Fig. \ref{fig-cmbpower} where the increase in  $\alpha$ presents the reduction of first acoustic peak in the CMB TT spectra. Thus, qualitatively the present unified model is slightly different from the GCG model. After that in Fig. \ref{fig-mpower}, we present the matter power spectra for the present unified model as well as for the GCG model where we also present a comparison between these unified models. In both the graphs of Fig. \ref{fig-mpower}, we have used different values of the key parameters $\mu$ and $\alpha$ of the corresponding models.  One can clearly see that with the increase of $\mu$, the matter power spectrum gets suppressed and this suppression is very transparent compared to 
the GCG model where we have considered a wide variation of the $\alpha$ parameter.   

\begin{figure*}
\includegraphics[width=0.35\textwidth]{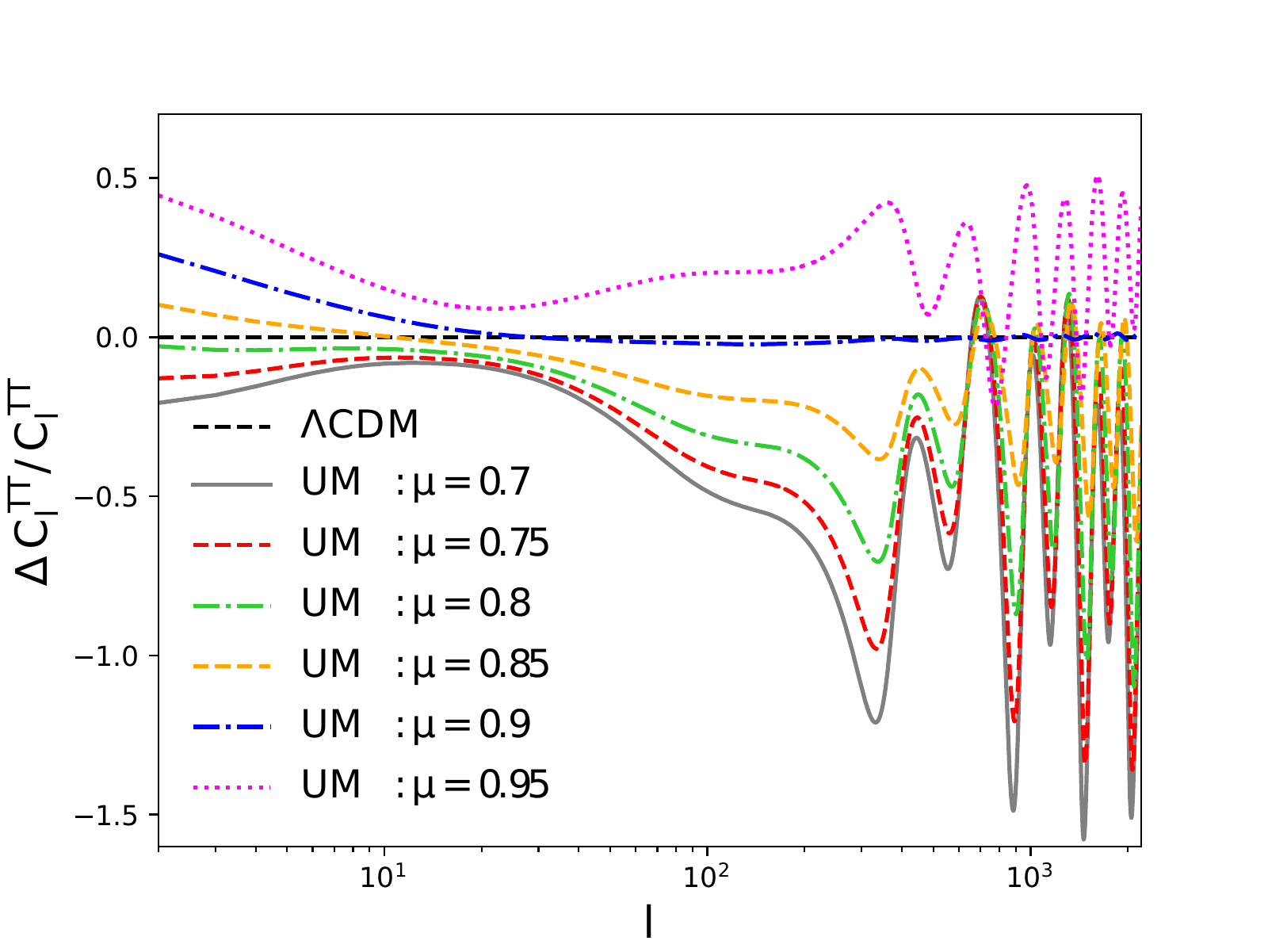}
\includegraphics[width=0.35\textwidth]{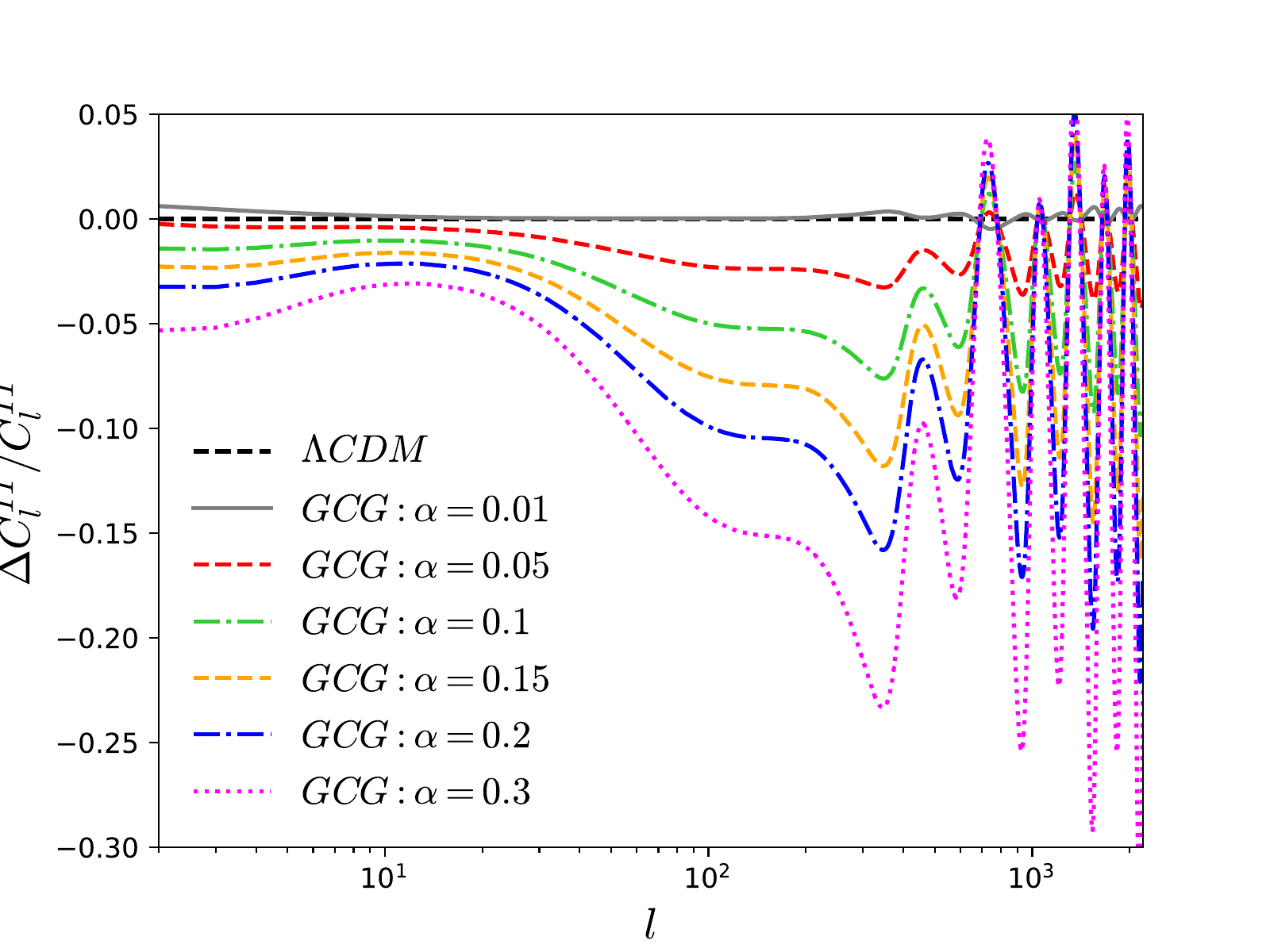}
\includegraphics[width=0.35\textwidth]{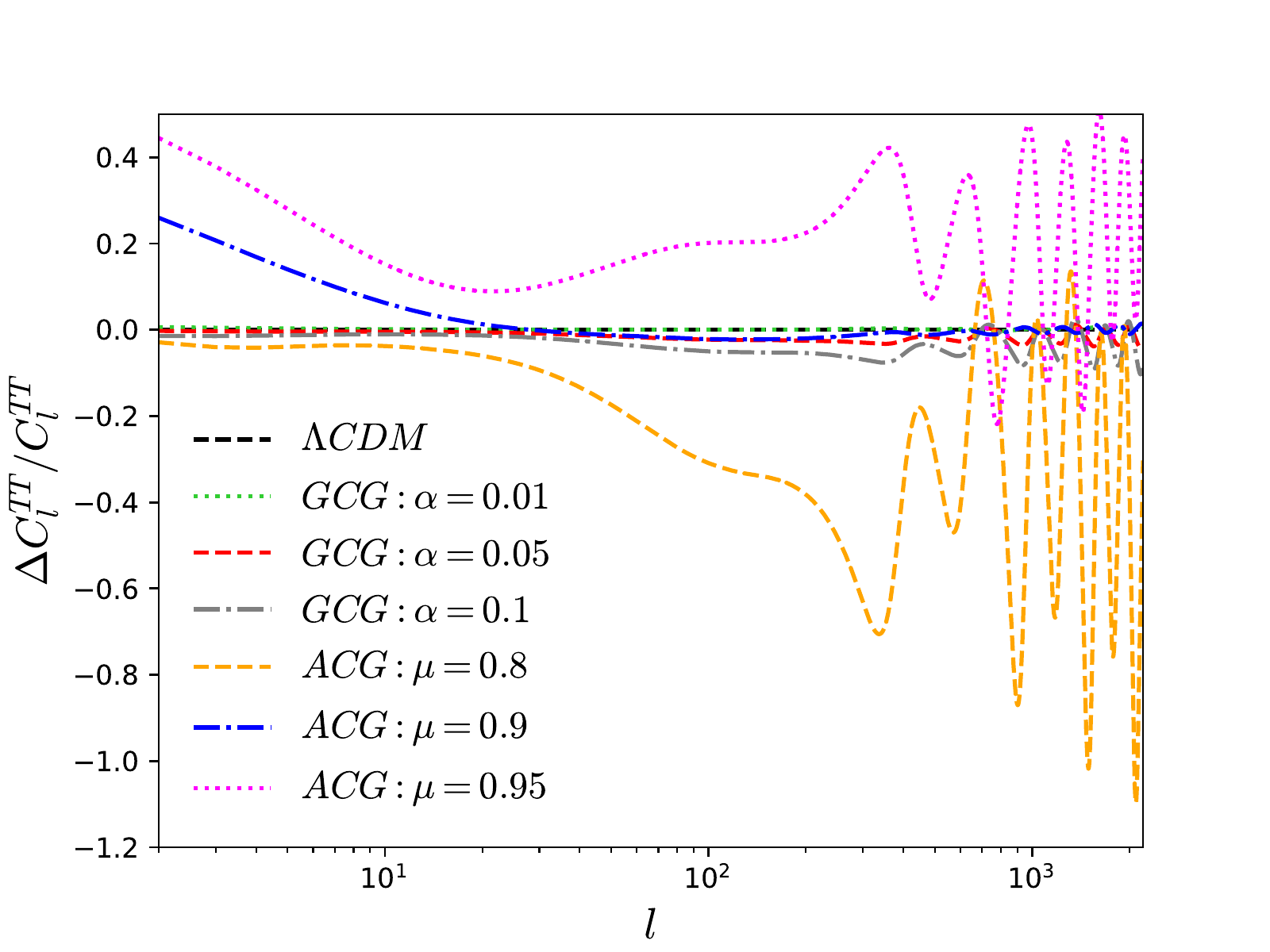}
\caption{In the upper left panel we show the relative deviation of the present unified model from the flat $\Lambda$CDM model using different values of $\mu$. The upper right panel shows the relative deviation of the GCG model with reference to the flat $\Lambda$CDM model. Finally, in the lower plot we include both the present unified and GCG models and compare them with respect to the reference model $\Lambda$CDM. }
\label{fig:ratio}
\end{figure*}

We close our discussion with Fig. \ref{fig:ratio} where we have shown three different plots for a better understanding of the present model in reference to GCG model as well as the $\Lambda$CDM. In the upper left panel of Fig. \ref{fig:ratio} we show the relative deviation of the present model with respect to the base model $\Lambda$CDM where we have taken several values of $\mu$ that correctly prescribe the transition from the decelerating to accelerating phase. We can see that the deviation of the present UM from the spatially flat $\Lambda$CDM model is clear while in the upper right panel of Fig. \ref{fig:ratio} we have shown the deviation of the GCG model with reference to the $\Lambda$CDM model with an aim to compare both the present UM and GCG models. From  the comparison between the left and right plots of the upper panel of Fig. \ref{fig:ratio}, one could realize that the deviation of the UM from the spatially flat $\Lambda$CDM is relatively higher compared to the deviation of GCG from $\Lambda$CDM. This feature has been made clear from an additional plot (see the lower plot of Fig. \ref{fig:ratio}) where we have considered both UM and GCG in a single frame and measured their deviations from the reference model $\Lambda$CDM.   
\begingroup                                                                                                                     
\squeezetable                                                                                                                   
\begin{center}                                                                                                                  
\begin{table*}                                                                                                                   
\begin{tabular}{ccccccccccccccccc}                                                                                                            
\hline\hline                                                                                                                    
Parameters & CMB+R18 & CMB+CC+R18 & CMB+Pantheon +R18 & CMB+Pantheon+CC+R18\\ \hline

$\Omega_b h^2$ & $    0.02211_{-    0.00014-    0.00027}^{+    0.00014+    0.00027}$ & $    0.02206_{-    0.00014-    0.00030}^{+    0.00014+    0.00029}$ & $    0.02194_{-    0.00015-    0.00030}^{+    0.00015+    0.00029}$ & $    0.02189_{-    0.00014-    0.00027}^{+    0.00014+    0.00028}$ \\

$100\theta_{MC}$ & $    1.02366_{-    0.00032-    0.00064}^{+    0.00034+    0.00066}$ & $    1.02374_{-    0.00033-    0.00065}^{+    0.00033+    0.00066}$ & $    1.02402_{-    0.00033-    0.00069}^{+    0.00032+    0.00067}$ & $    1.02406_{-    0.00032-    0.00066}^{+    0.00033+    0.00066}$  \\

$\tau$ & $    0.069_{-    0.016-    0.030}^{+    0.016+    0.032}$ & $    0.065_{-    0.016-    0.030}^{+    0.016+    0.032}$ & $    0.057_{-    0.016-    0.032}^{+    0.017+    0.033}$ & $    0.055_{-    0.017-    0.033}^{+    0.017+    0.034}$ \\

$n_s$ & $    0.9624_{-    0.0043-    0.0083}^{+    0.0042+    0.0081}$ & $    0.9607_{-    0.0044-    0.0080}^{+    0.0043+    0.0082}$ & $    0.9563_{-    0.0046-    0.0088}^{+    0.0048+    0.0085}$ & $    0.9550_{-    0.0039-    0.0078}^{+    0.0039+    0.0078}$ \\

${\rm{ln}}(10^{10} A_s)$ & $    3.077_{-    0.032-    0.061}^{+    0.033+    0.063}$ & $    3.070_{-    0.032-    0.061}^{+    0.032+    0.063}$ & $    3.058_{-    0.032-    0.063}^{+    0.033+    0.064}$ & $    3.055_{-    0.033-    0.067}^{+    0.033+    0.066}$ & \\

$\mu$ & $    0.905_{-    0.0028-    0.0057}^{+    0.0030+    0.0054}$ & $    0.903_{-    0.0030-    0.0058}^{+    0.0031+    0.0056}$ & $    0.899_{-    0.0034-    0.0070}^{+    0.0035+    0.0065}$ & $    0.897_{-    0.0031-    0.0059}^{+    0.0029+    0.0058}$ \\

$H_0$ & $   76.64_{-    0.65-    1.25}^{+    0.68+    1.25}$ & $   76.31_{-    0.67-    1.27}^{+    0.69+    1.26}$ & $   75.38_{-    0.72-    1.46}^{+    0.74+    1.41}$ & $   75.09_{-    0.65-    1.20}^{+    0.59+    1.25}$  \\

\hline 

$\chi^2_{\rm best-fit}$ & 12972.054 & 12989.536 & 14072.352 & 14071.216\\

\hline\hline                                                                                                                    
\end{tabular}                                                                                                                   
\caption{Here we have shown the 68\% and 95\% confidence-level constraints on the cosmological parameters of the present unified model after the addition of R18 data \cite{R18} to other cosmological  datasets. }
\label{tab:results2}                                                                                                   
\end{table*}                                                                                                                     
\end{center}                                                                                                                    
\endgroup 
\begin{table} 
\begin{tabular}{ccc}                
\hline\hline

$\ln B_{ij}$ & ~~~~~Strength of evidence for model ${M}_i$ \\ \hline
$0 \leq \ln B_{ij} < 1$ & Weak \\
$1 \leq \ln B_{ij} < 3$ & Definite/Positive \\
$3 \leq \ln B_{ij} < 5$ & Strong \\
$\ln B_{ij} \geq 5$ & Very strong \\
\hline\hline                        
\end{tabular}                                                                             
                             
\caption{Revised Jeffreys scale quantifying the statistical soundness of the models through the Bayesian evidence values \cite{Kass:1995loi}. } \label{tab:jeffreys}       
\end{table}

\begin{table*}
\small
\begin{center}                    
\begin{tabular}{ccccccccc}                                      \hline\hline              
 
Dataset & $\ln B_{ij}$ & Strength of evidence for $\Lambda$CDM \\ 
\hline

CMB &  $1.6$ & Definite/Positive \\

CMB+CC & $2.7$ & Definite/Positive \\

CMB+Pantheon & $4.5$ & Strong \\

CMB+Pantheon+CC & $3.6$ & Strong \\

\hline 

CMB+R18 &  $0.5$ & Weak \\

CMB+CC+R18 & $1.1$ & Definite/Positive\\

CMB+Pantheon+R18 & $2.9$ & Definite/Positive\\

CMB+Pantheon+CC+R18  & $2.3$ & Definite/Positive\\
\hline 
CC & $3.7$ & Strong \\
Pantheon & $7.9$ & Very Strong  \\
Pantheon+CC & $5.3$ & Very Strong \\
\hline\hline
\end{tabular}    
\caption{We display the values of $\ln B_{ij} = \ln B_i - \ln B_j$ (here $i$ stands for the reference model $\Lambda$CDM and $j$ stands for the unified model). From the table one can see that for all the combinations, $\Lambda$CDM is favored over the unified dark fluid model. } 
\label{tab:bayesian}                          
\end{center}    
\end{table*} 

\subsection{Bayesian Evidence}
\label{subsec-BE}

In this section we compute the Bayesian evidences of the present unified dark fluid model  aiming to understand the observational soundness of this model with respect to some reference cosmological model. Since  $\Lambda$CDM, is undoubtedly the best cosmological model at present, thus, we choose $\Lambda$CDM as the reference model. To compute the Bayesian evidences we use the publicly available code 
 \texttt{MCEvidence} \cite{Heavens:2017hkr,Heavens:2017afc} which directly takes the markov chain monte carlo chains of different observational analyses, such as CMB, CMB+CC, CMB+Pantheon and CMB+Pantheon+CC. For more discussions on the Bayesian evidence analysis, we refer to Ref. \cite{Yang:2018qmz}. In
Table~\ref{tab:jeffreys} we show the  revised Jeffreys scale \cite{Kass:1995loi}  .   
quantifying the statistical comparison of the models through the Bayesian evidence values.

Now, using the \texttt{MCEvidence} \cite{Heavens:2017hkr,Heavens:2017afc} we calculate the $\ln B_{ij}$ values for the unified dark fluid model for all the observational datasets including the background datasets as well. In Table \ref{tab:bayesian} we sumamrize the values of  $\ln B_{ij}$ where we refer $i$ for the reference model $\Lambda$CDM and $j$ for the present unified model. From Table \ref{tab:bayesian}, one can see that for all the datasets (including CMB) $\Lambda$CDM is favored over the unfiied model. Once again we see that for the background data only, the evidence in favor of $\Lambda$CDM increases, which is suppressed a bit when the CMB data are added to them. In summary, from point of view of the Bayesian evidence analyses,  $\Lambda$CDM is still a preferred candidate for the universe's evolution.

\section{Summary and conclusions}
\label{sec-conclu}

The dynamics of our universe as reported by various observational data is very complicated. Almost 96\% of the total energy budget of the universe is occupied by some dark sector where the maximum percentage $\sim 68\%$ comes from some hypothetical dark energy fluid and around $28\%$ is coming from some non-luminous dark matter species. The origin  origin, nature and evolution of both these dark fluids are absolutely unknown to us. Thus, modelling the universe's evolution has been a great deal for modern cosmology and various attempts have been made so far. Since the overall identity of these dark fluids is not clear, thus, different approaches are used to depict the expansion history of the universe. One of the commonly used mechanisms is to consider that
dark matter and dark energy evolve separately, that means, the dynamics of each dark fluid is independent of the other. While it is also conjectured that both these dark fluids are actually two different faces of a single fluid that actually is playing the role of both the dark fluids. 
The proposal of some single dark exhibiting two different dark sides of the universe is commonly known as the  {\it unified dark fluid} in the cosmological literature. 
In the present work we have focused  on unified dark fluid model. 
The Chaplygin gas is a renowned  name in this context where a series of Chaplygin type models, starting from the simple  Chaplygin gas to modified Chaplygin model  have been introduced and investigated with the observational data. Although Chaplygin models have got much attention in the cosmological community, however,  concerning the observational issues, the Chaplygin models are diagnosed with some problems \cite{Sandvik:2002jz,Li:2009mf}. So,  a natural attempt could be to widen the allowance of other unified cosmological models in order to test their acceptability with the recent observational data. Thus, in the present work we investigate a specific unified cosmological model that was introduced in \cite{Hova:2010na} and recently investigated in \cite{Hernandez-Almada:2018osh}. 
The model is quantified by only a single parameter $\mu$, similar to the GCG model $p =A/\rho^{\alpha}$ ($A > 0$, $ 0 \geq \alpha 1$) having only one free parameter $\alpha$. 
In both the articles, namely,  \cite{Hova:2010na} and \cite{Hernandez-Almada:2018osh}, the authors discussed the evolution of the model at the level of background, however no such analysis at the level of perturbations was performed. With the feeling and experience with the cosmological models at the level of perturbations, we have performed a robust statistical fittings of the model with a special focus on its behaviour in the large scale structure of the universe. 

We started with the theoretical analysis of the model where we have investigated various cosmological parameters and compared the model with the known unified model GCG. In Fig. \ref{fig-eos}, Fig. \ref{fig-deceleration} and Fig. \ref{fig-Omega} we have studied various important cosmological parameters. From the evolution of the deceleration parameter for this model (left panel of Fig. \ref{fig-deceleration}), we found a restriction on the key parameter $\mu$ of the present model  through the transition of the universe from its past decelerating phase to the current accelerating phase. 

We then perform a robust observational analysis with this unified model by extracting its cosmological constraints using various cosmological data presented in Table \ref{tab:results1} and \ref{tab:results2}. Concerning the key parameter of the unified model, namely, $\mu$, we find that, $\mu \sim 0.9$, for all the observational datasets employed in this work. Consequently, the above estimation of $\mu$ implies that $q \sim -0.8$ (see the left panel of Fig. \ref{fig-deceleration}) which is in tension with a recent model independent estimation of the deceleration parameter, $q = -0.52 \pm 0.06$ \cite{Haridasu:2018gqm}. So, the present allows a super accelerating phase of the universe \cite{Kaplinghat:2003vf,Cai:2005qm,Das:2005yj,Kaplinghat:2006jk}. 
From the analyses, we found that irrespective of the present datasets, we always have a higher $H_0$, which is even slightly higher than the local estimation by Riess \cite{Riess:2016jrr,R18,Riess:2019cxk}, thus, eventually, one could realize that the tension on $H_0$ is alleviated. This is one of the interesting results of this work and might be considered as a potential model in the list of cosmological models alleviating the $H_0$ tension.  

We then investigated the behaviour of  this model at large scale via CMB TT and matter power spectra.  We find that this unified model 
is surely deviating from the GCG and $\Lambda$CDM model, but the qualitative features of the model remains same to that of the GCG model. This might be a weak point of this model since the Chaplygin type models are diagnosed with some problems as quoted in \cite{Sandvik:2002jz,Li:2009mf}.  
However, from the results of our analysis we can see that unified models which are extensions of the GCG cannot be  excluded from the cosmological data. However, from the Bayesian evidence analysis, $\Lambda$CDM is always favored over the unified dark fluid model.  Further analyses with this model are necessary in order to gain more visibility on this model.

Last but not least, we would like to comment that since the present unified model is new in the cosmological society, and so far we are aware of the literature, not enough investigations are performed with this model. Thus, based on the present result, mainly on its potentiality to alleviate the $H_0$ tension, we believe that it will be interesting to study this model further using future observational data from different astronomical missions, for example, Gravitational waves standard sirens from different observatories  \cite{Punturo:2010zz,Audley:2017drz,Kawamura:2011zz,Luo:2015ght}, and some other astronomical missions like Simons Observatory Collaboration \cite{Ade:2018sbj}, Cosmic Microwave Background Stage-4 \cite{Abitbol:2017nao}, EUCLID Collaboration \cite{Laureijs:2011gra},  Dark Energy Spectroscopic Instrument \cite{Aghamousa:2016zmz}, etc. 
The effects of neutrinos could be another direction of research in this context. Such analyses are certainly interesting and indeed open to all. We plan to report some of them in near future.

\section{Acknowledgments}
The authors thank the referee for several essential comments that improved the the manuscript substantially. WY has been supported by the National Natural Science Foundation of China under Grants No. 11705079 and No. 11647153. SP acknowledges the partial support from the Faculty Research and Professional Development Fund (FRPDF) Scheme  of Presidency University, Kolkata, India.  YW has been supported by the National Natural Science Foundation of China under Grant No. 11575075.


\end{document}